\documentclass[iop,revtex4]{emulateapj} 

\usepackage{graphicx,subfigure}
\usepackage{float}
\usepackage{booktabs}
\usepackage{color}

\newcommand{\comm}[1]{}

\shorttitle{The Subaru Coronagraphic Extreme Adaptive Optics system}
\shortauthors{N. Jovanovic et al.}

\begin{document}
\title{The Subaru Coronagraphic Extreme Adaptive Optics system: \\ enabling high-contrast imaging on solar-system scales}
\author{N. Jovanovic\altaffilmark{1}, F. Martinache\altaffilmark{2}, O. Guyon\altaffilmark{1,3,4}, C. Clergeon\altaffilmark{1}, G. Singh\altaffilmark{1,5}, T. Kudo\altaffilmark{1},V. Garrel\altaffilmark{6}, \\
K. Newman\altaffilmark{4,7},  D. Doughty\altaffilmark{1,4}, J. Lozi\altaffilmark{1}, J. Males\altaffilmark{3,8}, Y. Minowa\altaffilmark{1}, Y. Hayano\altaffilmark{1}, N. Takato\altaffilmark{1}, J. Morino\altaffilmark{9}, \\
J. Kuhn\altaffilmark{10}, E. Serabyn\altaffilmark{10}, B. Norris\altaffilmark{11}, P. Tuthill\altaffilmark{11}, G. Schworer\altaffilmark{5,11}, P. Stewart\altaffilmark{11}, L. Close\altaffilmark{3}, E. Huby\altaffilmark{5,12}, \\ G. Perrin\altaffilmark{5},  S. Lacour\altaffilmark{5}, L. Gauchet\altaffilmark{5}, S. Vievard\altaffilmark{5}, N. Murakami\altaffilmark{13}, F. Oshiyama\altaffilmark{13}, N, Baba\altaffilmark{13}, T. Matsuo\altaffilmark{14}, \\
J. Nishikawa\altaffilmark{9}, M. Tamura\altaffilmark{9,15}, O. Lai\altaffilmark{1,6}, F. Marchis\altaffilmark{16}, G. Duchene\altaffilmark{17,18}, T. Kotani\altaffilmark{9}, and J. Woillez\altaffilmark{19}}
\altaffiltext{1}{National Astronomical Observatory of Japan, Subaru Telescope, 650 North A'Ohoku Place, Hilo, HI, 96720, U.S.A.}
\altaffiltext{2}{Observatoire de la Cote d'Azur, Boulevard de l'Observatoire, Nice, 06304, France}
\altaffiltext{3}{Steward Observatory, University of Arizona, Tucson, AZ, 85721, U.S.A.}
\altaffiltext{4}{College of Optical Sciences, University of Arizona, Tucson, AZ 85721, USA}
\altaffiltext{5}{LESIA, Observatoire de Paris, Meudon, 5 Place Jules Janssen, 92195, France.}
\altaffiltext{6}{Gemini Observatory, c/o AURA, Casilla 603, La Serena, Chile}
\altaffiltext{7}{NASA Ames Research Center, Moffett Field, CA 94035, USA}
\altaffiltext{8}{NASA Sagan Fellow}
\altaffiltext{9}{National Astronomical Observatory of Japan, 2-21-1 Osawa, Mitaka, Japan}
\altaffiltext{10}{Jet Propulsion Laboratory, California Institute of Technology, 4800 Oak Grove Dr, Pasadena, CA 91109, USA}
\altaffiltext{11}{Sydney Institute for Astronomy (SIfA), Institute for Photonics and Optical Science (IPOS), School of Physics, \\University of Sydney, NSW 2006, Australia}
\altaffiltext{12}{Département d?Astrophysique, Géophysique et Océanographie, Université de Liège, 17 Allée du Six Août, 4000 Liège, Belgium}
\altaffiltext{13}{Division of Applied Physics, Faculty of Engineering, Hokkaido University, Kita-13, Nishi-8, Kita-ku, Sapporo, Hokkaido 060-8628, Japan}
\altaffiltext{14}{Kyoto University, Kitashirakawa-Oiwakecho, Sakyo-ku, Kyoto 606-8502 Japan}
\altaffiltext{15}{Department of Astronomy, University of Tokyo, 7-3-1 Hongo, Bunkyo, Tokyo 113-0033, Japan}
\altaffiltext{16}{Carl Sagan Center at the SETI Institute, Mountain View, CA 94043, USA}
\altaffiltext{17}{Astronomy Department, University of California, Berkeley, CA 94720-3411, USA}
\altaffiltext{18}{University Grenoble Alpes \& CNRS, Institut de Planetologie et d'Astrophysique de Grenoble (IPAG), Grenoble F-3800, France}
\altaffiltext{19}{European Southern Observatory (ESO), Karl-Schwarzschild-Str. 2, Garching 85748, Germany}
\email{jovanovic.nem@gmail.com}

\begin{abstract}
The Subaru Coronagraphic Extreme Adaptive Optics (SCExAO) instrument is a multipurpose high-contrast imaging platform designed for the discovery and detailed characterization of exoplanetary systems and serves as a testbed for high-contrast imaging technologies for ELTs. It is a multi-band instrument which makes use of light from $600$ to $2500$~nm allowing for coronagraphic direct exoplanet imaging of the inner $3~\lambda/D$ from the stellar host. Wavefront sensing and control are key to the operation of SCExAO. A partial correction of low-order modes is provided by Subaru's facility adaptive optics system with the final correction, including high-order modes, implemented downstream by a combination of a visible pyramid wavefront sensor and a $2000$-element deformable mirror. The well corrected NIR (y-K bands) wavefronts can then be injected into any of the available coronagraphs, including but not limited to the phase induced amplitude apodization and the vector vortex coronagraphs, both of which offer an inner working angle as low as $1~\lambda/D$. Non-common path, low-order aberrations are sensed with a coronagraphic low-order wavefront sensor in the infrared (IR). Low noise, high frame rate, NIR detectors allow for active speckle nulling and coherent differential imaging, while the HAWAII $2$RG detector in the HiCIAO imager and/or the CHARIS integral field spectrograph (from mid $2016$) can take deeper exposures and/or perform angular, spectral and polarimetric differential imaging. Science in the visible is provided by two interferometric modules: VAMPIRES and FIRST, which enable sub-diffraction limited imaging in the visible region with polarimetric and spectroscopic capabilities respectively. We describe the instrument in detail and present preliminary results both on-sky and in the laboratory. 
\end{abstract}


\section{Introduction}
The field of high-contrast imaging is advancing at a great rate with several extreme adaptive optics systems having come online in $2014$, including the Gemini Planet Imager (GPI) \citep{macintosh14}, the Spectro-Polarimetric High-contrast Exoplanet REsearch instrument (SPHERE) \citep{bez08}, and the focus of this work, the Subaru Coronagraphic Extreme Adaptive Optics (SCExAO) system which join the already running P$1640$~\citep{dekany13}. These systems all share a similar underlying architecture: they employ a high order wavefront sensor (WFS) and a deformable mirror (DM) to correct for atmospheric perturbations enabling high Strehl ratios in the near-infrared (NIR) ($>90\%$), while a coronagraph is used to suppress on-axis starlight downstream. The primary motivation for such instrumentation is the direct detection of planetary mass companions at contrasts of $10^{-5}$--$10^{-6}$ with respect to the host star, at small angular separations (down to $1$--$5~\lambda/D$) from the host star. 

The era of exoplanetary detection has resulted in $\sim1500$ planets so far confirmed~\citep{han14}. The majority of these were detected via the transit technique with instruments such as the Kepler space telescope~\citep{borucki10}. The radial velocity method~\citep{mayor95} has also been prolific in detection yield. Both techniques are indirect in nature (the presence of the planet is inferred by its effect on light from the host star) and hence often deliver limited information about the planets themselves. It has been shown that it is possible to glean insights into atmospheric compositions via techniques such as transit spectroscopy \citep{char01}, whereby star light from the host passes through the upper atmosphere of the planet as it propagates to Earth, albeit with limited signal-to-noise ratio. The ability to directly image planetary systems and conduct detailed spectroscopic analysis is the next step towards understanding the physical characteristics of their members and refining planetary formation models.

To this end, so far $<50$ substellar companions have been directly imaged (see Fig.~3 in ~\citep{pepe14}). The challenge lies in being able to see a companion, many orders of magnitude fainter, at very small angular separations ($<1"$), from the blinding glare of the host star. Indeed, the Earth would be $>10^{9}\times$ fainter than the sun if viewed from outside the solar system in reflected light. Although these levels of contrast can not be overcome from ground based observations at small angular separations ($<0.5"$), it is possible to circumvent this by imaging the thermal signatures instead and targeting bigger objects. Indeed, all planets imaged thus far were large Jupiter-like planets (which are brightest) detected at longer wavelengths (in the near-IR H and K-bands and the mid-IR L and M-bands) in thermal light (a subset of detections include~\citep{marois08,Lag09,kraus12}). To overcome the glare from the star which results in stellar photons swamping the signal from the companion, adaptive optics systems (AO) are key~\citep{Lag09}. Although angular differential imaging is the most commonly used technique for imaging planets thus far~\citep{marois08}, coronagraphy~\citep{Laf09,serebyn2010a} and aperture-masking interferometry~\citep{kraus12} have also been used to make detections. With the direct detection of the light from the faint companion itself, spectroscopy becomes a possibility and indeed preliminary spectra have been taken for some objects as well~\citep{bar11,Opp13}.

In addition to planetary spectroscopy, how disks evolve to form planetary systems is a key question that remains unanswered. Thus far coronographic imagers like HiCIAO at the Subaru Telescope have revealed intricate features of the inner parts of circumstellar disks using polarization differential imaging (under the SEEDS project~\citep{tamura09}). These solar-system scale features include knots and spiral density waves within disks like MWC$758$ and SAO $206462$ \citep{grady13,muto12}. How such features are affected by or lead to the formation and evolution of planets can only be addressed by high-contrast imaging of the inner parts (up to $15$~AU from the star) of such disks. To address the lack of information in this region, high-contrast imaging platforms equipped with advanced wavefront control and coronagraphs are pushing for smaller inner working angles (IWA). In the limit of low wavefront aberrations, currently achieved with AO systems operating in the near-IR, coronagraphs are the ideal tool for imaging the surrounding structure/detail as they are unrivaled in achievable contrast. Both the contrast and IWA are dependent on the level of wavefront correction available. With wavefront corrections typically offered by facility adaptive optics (AO) systems on $5$--$10$~m class telescopes ($\sim30-40\%$ in H-band) previous generations of coronagraphic imagers, such as the Near-Infrared Coronagraphic Imager, NICI on the Gemini South telescope~\citep{artigau08} were optimized for an IWA of $5-10~\lambda/D$. However, with extreme AO (ExAO) correction offering high Strehl ratio and stable pointing, GPI and SPHERE have been optimized for imaging companions down to angular separations of $\sim3~\lambda/D$ ($>120$~mas in the H-band). SCExAO utilizes several more sophisticated coronagraphs including the Phase Induced Amplitude Apodization (PIAA)~\citep{guyon03} and vector vortex~\citep{mawet10}, which drive the IWA down to just below $1 \lambda/D$. At a distance of $100$~pc, the PIAA/vortex coronagraphs on SCExAO would be able to image from $4$~AU outwards (approximately the region beyond the orbit of Jupiter). Further, in the case of the HR$8799$ system the IWA would be $1.6$~AU (the distance of Mars to our Sun) making it possible for SCExAO to image the recently hypothesized 5\textsuperscript{th} planet at $9$~AU (mass between $1$-$9$ Jupiter masses)~\citep{goz14} if it is indeed present as predicted. 

Despite the state-of-the-art IWA offered by these coronagraphs in the near-IR, the structure of disks and the distribution of planets at even closer separations than $1~\lambda/D$ will remain inaccessible with coronagraph technology alone. This scale is scientifically very interesting as it corresponds to the inner parts of the solar system where the majority of exoplanets have been found to date based on transit and radial velocity data~\citep{han14}. To push into this regime SCExAO uses two visible wavelength interferometric imaging modules known as VAMPIRES~\citep{norris12a} and FIRST~\citep{huby12}. VAMPIRES is based on the powerful technique of aperture masking interferometry \citep{tuthill00}, while FIRST is based on an augmentation of that technique, known as pupil remapping~\citep{perrin06}. Operating in the visible part of the spectrum, the angular resolution of these instruments on an $8$-m class telescope approaches a territory previously reserved for long baseline interferometers ($15$~mas at $\lambda=700$~nm) and expands the type of target that can be observed to include massive stars. Although the modules operate at shorter wavelengths where the wavefront correction is of a lower quality, interferometric techniques allow for sub-diffraction limited imaging ($10$~mas resolution using $\lambda/2D$ criteria as conventionally used in interferometry) even in this regime, albeit at lower contrasts ($\sim10^{-3}$), making optimal use of the otherwise discarded visible light. Despite the lower contrasts aperture masking interferometry has already delivered faint companion detections at unprecedented spatial scales~\citep{kraus12}. Each module additionally offers a unique capability. For example the polarimetric mode of VAMPIRES is designed to probe the polarized signal from dusty structures such as disks around young stars and shells around giant stars \citep{norris12b,norris15} at a waveband where the signal is strongest. This is a visible analog of that offered by the SAMPol mode on the NACO instrument at the VLT~\citep{Lenzen03, tuthill10, norris12b}. FIRST on the other hand offers the potential for broadband spectroscopy and is tailored to imaging binary systems and the surface features of large stars. Such capabilities greatly extend SCExAO beyond that of a regular ex-AO facility.

Finally, with a diffraction-limited point-spread-function (PSF) in the near-IR, and a large collecting area, SCExAO is ideal for injecting light into fiber fed spectrographs such as the Infrared Doppler instrument (IRD)~\citep{tamura12}. In addition, this forms the ideal platform for exploring photonic-based technologies such as photonic lanterns~\citep{saval2013} and integrated photonic spectrographs~\citep{cvetojevic12} for next generation instrumentation. 

The aim of this publication is to outline the SCExAO instrument and its capabilities in detail and offer some preliminary results produced by the system. In this vein, section~\ref{sec:elements} describes the key components of SCExAO while section~\ref{sec:fund} highlights the functionalities and limitations of the instrument. Section~\ref{sec:future} outlines plans for future upgrades and the paper concludes with a summary in Section~\ref{sec:summary}.

\section{The elements of SCExAO}\label{sec:elements}
\begin{figure*}
\centering 
\includegraphics[width=0.75\linewidth]{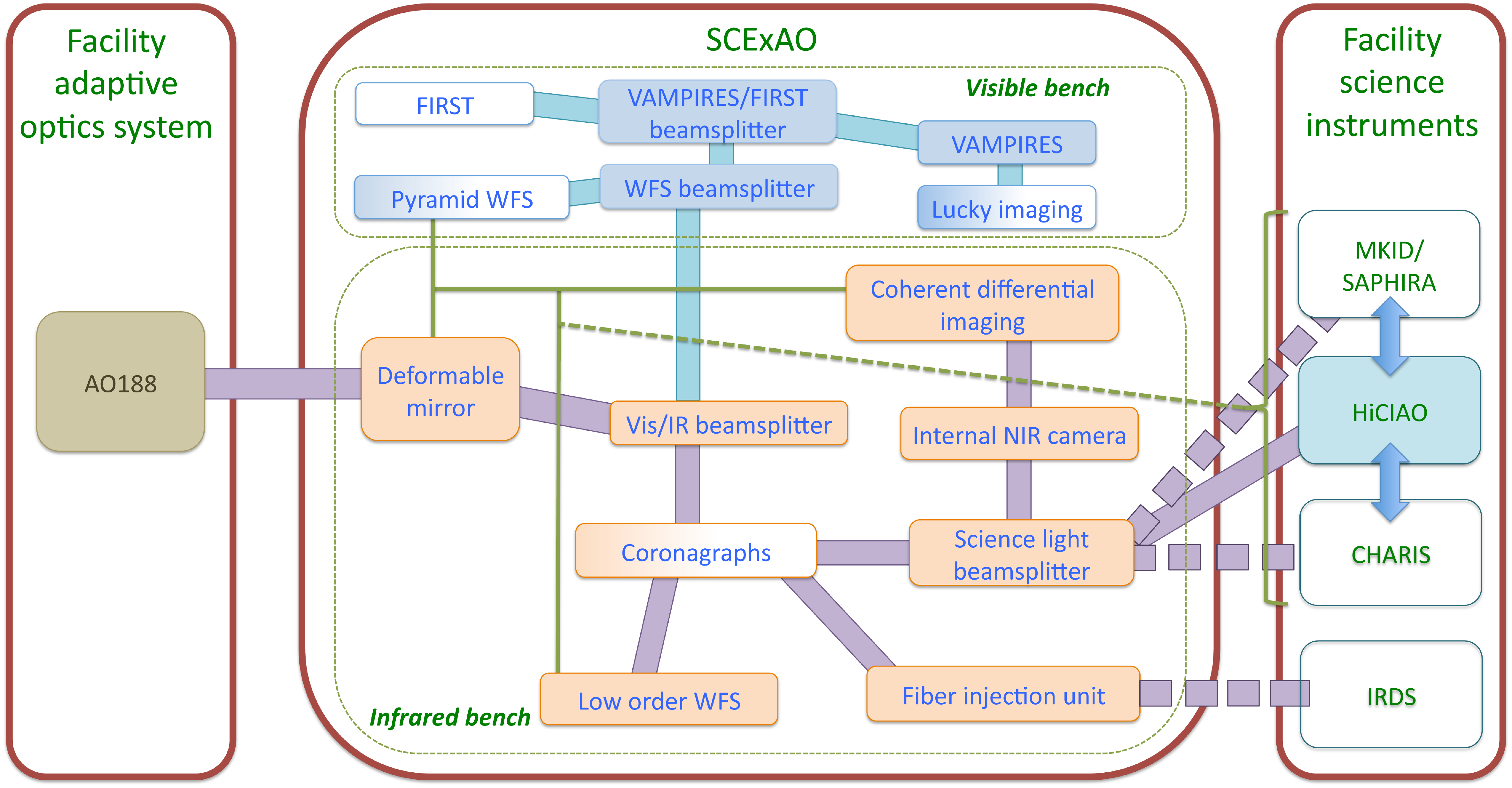}
\caption{\footnotesize System level flow diagram of the SCExAO instrument. Thick purple and blue lines depict optical paths while thin green lines signify communication channels. Dashed lines indicate that a connection does not currently exist but it is planned for the future. Box fill indicates commissioning status. Solid boxes: commissioned, graded boxes: partially commissioned, white background: not commissioned.}
\label{fig:scexao_sl}
\end{figure*}

\begin{figure*}
\centering 
\includegraphics[width=0.75\linewidth]{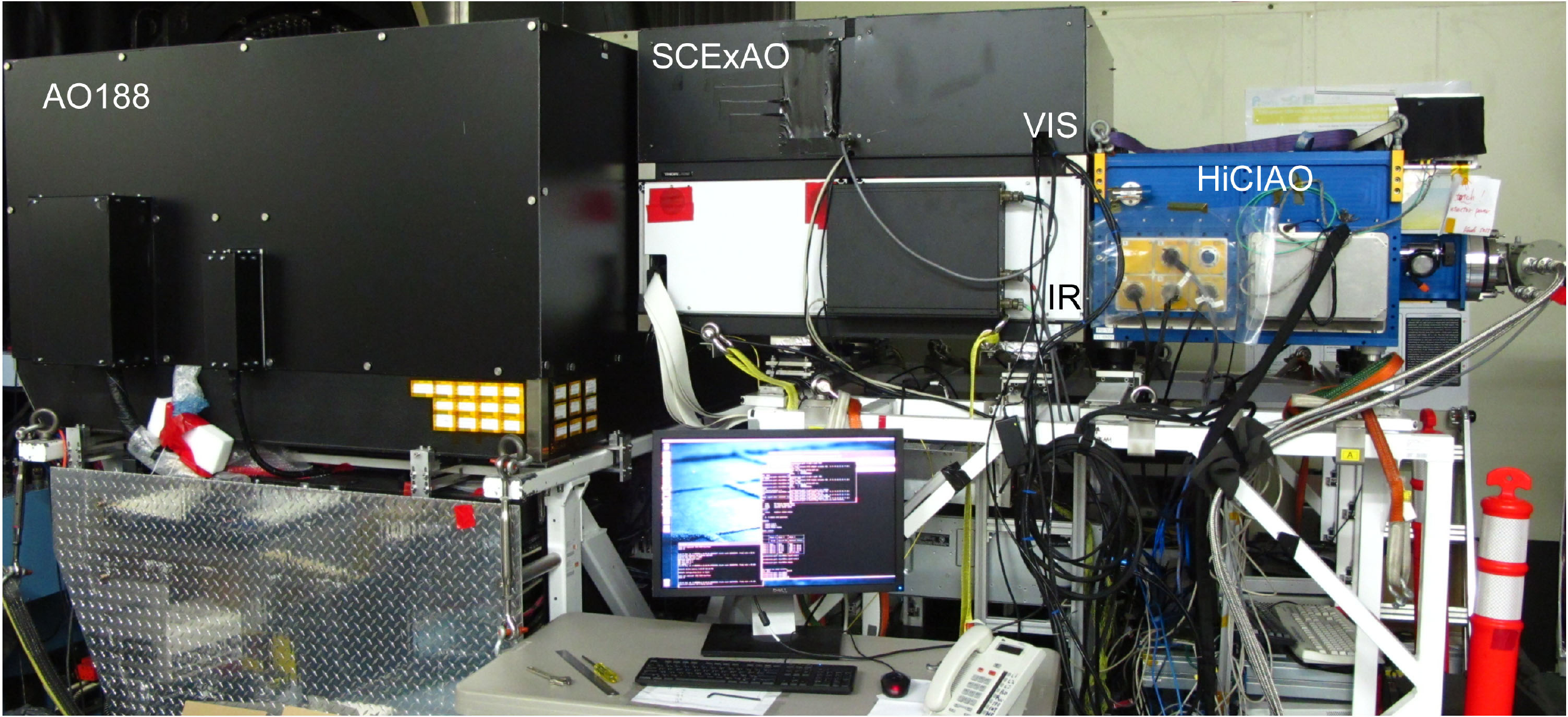}
\caption{\footnotesize Image of SCExAO mounted at the Nasmyth IR platform at Subaru Telescope. To the left is AO$188$ which injects the light into SCExAO (center) and HiCIAO is shown on the right. The FIRST recombination bench is not shown for visual clarity.}
\label{fig:scexao_nas}
\end{figure*}

In order to understand the scientific possibilities and limitations of the SCExAO instrument, it is important to first understand the components and their functionalities. To aid the discussion a system level diagram of SCExAO is shown in Fig.~\ref{fig:scexao_sl}, and an image of the instrument at the Nasmyth platform is shown in Fig.~\ref{fig:scexao_nas}. A detailed schematic of the major components is shown in Fig.~\ref{fig:scexao}. The components and functionalities of SCExAO have been undergoing commissioning as will be outlined throughout this publication. A summary of the commissioning status of each mode of operation or module of SCExAO can be found in Table~\ref{tab:commission}.

The main aim of SCExAO is to exploit the well corrected wavefront enabled by the high order WFS to do high-contrast imaging with light across a broad spectrum: from $600$~nm to $2500$~nm. As such there are a number of instrument modules within SCExAO that operate in different wavebands simultaneously while the coronagraph is collecting data. This hitchhiking mode of operation enables maximum utilization of the stellar flux, which allows for a more comprehensive study of each target. 

 
\subsection{SCExAO at a glance}
\begin{figure*}
\centering 
\includegraphics[width=0.99\linewidth]{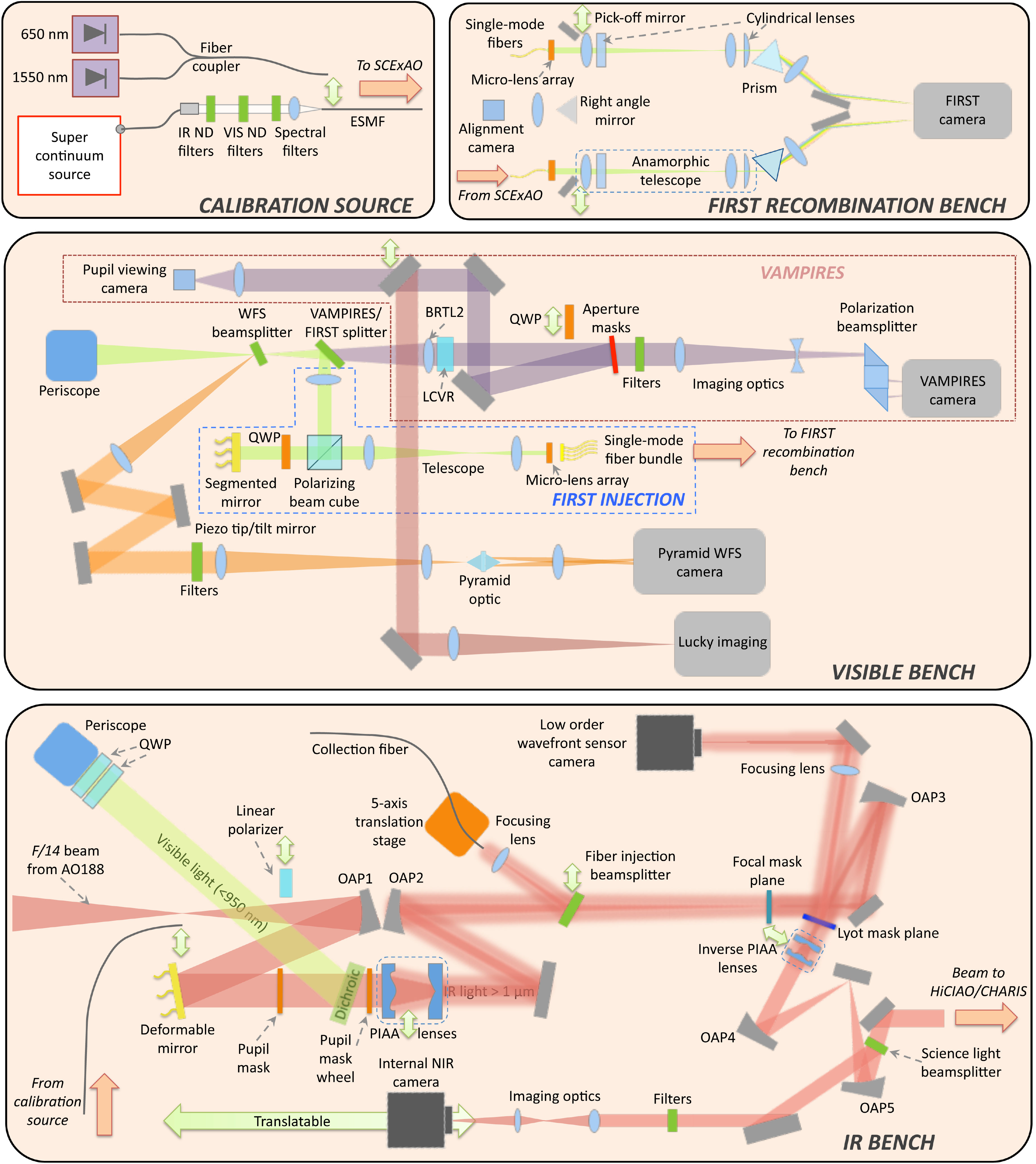}
\caption{\footnotesize Schematic diagram of the SCExAO instrument. Top box (left): Portable calibration source layout. Top box (right): FIRST recombination bench. Middle box: layout of the visible optical bench which is mounted on top of the IR bench. Bottom box: IR bench layout. Dual head green arrows indicate that a given optic can be translated in/out of or along the beam. Orange arrows indicate light entering or leaving the designated bench at that location.}
\label{fig:scexao}
\end{figure*}

The SCExAO instrument consists of two optical benches mounted on top of one another, separated by $\sim350$~mm. The bottom bench (IR bench) hosts the deformable mirror, coronagraphs, and a Lyot-based low order wavefront sensor (LLOWFS) while the top bench (visible bench) hosts the pyramid WFS, VAMPIRES, FIRST and lucky imaging (see Figure~\ref{fig:scexao}). The benches are optically connected via a periscope.

The light from the facility adaptive optics system (AO$188$) is injected into the IR bench of SCExAO and is incident on the $2000$ element deformable mirror (2k DM) before it is split by a dichroic into two distinct channels: light shorter than $940$~nm is reflected up the periscope and onto the top bench while light longer than $940$~nm is transmitted. The visible light is then split by spectral content by a range of long and short pass dichroics which send the light to the pyramid WFS (PyWFS) and visible light science instruments. The PyWFS is used for the high order wavefront correction and drives the DM on the IR bench. The VAMPIRES and FIRST modules utilize the light not used by the PyWFS. Lucky imaging/PSF viewing makes use of light rejected by the aperture masks of VAMPIRES.

The IR light that is transmitted by the dichroic on the IR bench propagates through one of the available coronagraphs. After the coronagraphs, the light reflected by the Lyot stop is used to drive a LLOWFS in order to correct for the chromatic and non-common errors (such as tip/tilt) between the visible and IR benches~\citep{singh14}. The light transmitted by the coronagraphs is then incident on the science light beamsplitter which determines the spectral content and exact amount of flux to be sent to a high frame rate internal NIR camera as compared to a science grade detector such as the HAWAII 2RG in the HiCIAO instrument and soon to be commissioned CHARIS. The internal NIR camera can then be used to drive various coherent differential imaging algorithms. 

\subsection{Detailed optical design}
The instrument is designed to receive partially corrected light from the facility adaptive optics system, AO$188$ ($188$ actuator deformable mirror). The beam delivered from AO$188$ converges with a speed of $f/14$. Typical H-band Strehl ratios are $\sim30-40\%$ in good seeing~\citep{minowa10}. The beam is collimated by an off-axis parabolic mirror (OAP$1$, $f=255$~mm) creating an $18$~mm beam. Details of the OAPs can be found in section~\ref{sec:appendix}. The reflected beam is incident upon the 2k DM, details of which are in section~\ref{sec:DM}. The surface of the DM is placed one focal length from OAP$1$ which conjugates it with the primary mirror of the telescope (i.e. it is in a pupil plane). Once the beam has reflected off the DM it is incident upon a fixed pupil mask which replicates the central obstruction and spiders of the telescope (see Fig.~\ref{fig:PSF}), albeit slightly oversized. This mask is permanently in the beam (both on-sky and in the laboratory) so that response matrices collected for the various wavefront sensors with the internal light source in SCExAO can be used on-sky as well. It is positioned as close to the pupil plane as possible ($\sim70$~mm away from the DM) and forms the primary pupil for the instrument. The image/pupil rotator in AO$188$ is used to align the telescope pupil with the fixed internal mask when on-sky (i.e. SCExAO operates with a fixed pupil and rotating image when observing). 

Immediately following the mask is a dichroic beamsplitter ($50$~mm diameter, $7$~mm thick) which reflects the visible light ($<940$~nm) and transmits the IR ($>940$~nm). In the transmitted beam path, there is a mask wheel after the dichroic which hosts numerous masks including the shaped pupil coronagraphic mask. The automated mounts for the phase induced amplitude apodization (PIAA) coronagraph lenses are adjacent to the mask wheel. They too were placed as close to the DM pupil as possible.  The PIAA lenses themselves will be described in detail, for now it is important to note that they can be retracted from the beam entirely. A flat mirror is used to steer the light onto OAP$2$ which focuses the beam ($f=519$~mm). A variety of coronagraphic focal plane masks, used to suppress the on-axis star light are housed in a wheel which also has three axes of translation in the focal plane (the masks are outlined in sections~\ref{sec:cor}). OAP$3$ recollimates the beam to a $9$~mm diameter beam ($f=255$~mm). A wheel with Lyot plane masks is situated in the collimated beam such that the masks are conjugated with the pupil plane. The Lyot wheel can be adjusted in lateral alignment via motorized actuators. Light diffracted by the focal plane mask and reflected from the Lyot stop is imaged onto a non-science grade (i.e. relatively high noise), high frame rate detector which is used for low order wavefront sensing. The light transmitted by the Lyot stop is next incident upon the inverse PIAA lenses (detailed in~\ref{sec:cor}). They are mounted on stages which allow motorized control of the lateral positioning and are conjugated to the PIAA lenses upstream and can also be retracted from the beam entirely. OAP$4$ and OAP$5$ reimage the telescope pupil for the HiCIAO/CHARIS camera. On its way to the camera the beam is intercepted by the science light beamsplitter which is used to control the flux and spectral content sent to a high frame rate internal NIR camera and the facility science instruments. Detectors in a high-contrast imaging instrument are amongst the most important components and therefore the performance of those used in SCExAO is summarized in Table~\ref{tab:detectors}. 

\begin{deluxetable*}{cccccccc}[h!]
\tabletypesize{\footnotesize}
\centering
\tablecaption{Detector characteristics used within SCExAO.  \label{tab:detectors}}
\tablehead{
\colhead{Detector}  			& 	\colhead{Technology	}	& 	\colhead{Detector size}  		&  \colhead{Pixel size}      	&    \colhead{Read-out}  	  	&   \colhead{Full frame}       	&  \colhead{Operating} 			&     \colhead{Manufacturer}	\\
\colhead{Name}				& 					&	     \colhead{(pixels)} 		&   	\colhead{($\mu$m)}		&   \colhead{noise ($e^{-}$)}  	&  \colhead{(Frame-rate (Hz))}		   	&  \colhead{Wavelengths (nm)}		&    \colhead{(Product name)}} 	\\  
\startdata
Internal    					&  	InGaAs 			& 	$320\times256$ 	&    $30$			&   $114$			&  $170$				&	$900-1700$		&  Axiom Optics	\\
NIR						&	(CMOS)			&					&				&				&  					&					& (OWL SW1.7HS)  	\\
camera					&					&					&				&				&					&					&				\\
LOWFS					&  	InGaAs 			&   	$320\times256$ 	&    $30$			&   $140$			&  $170$				&      $900-1700$		& Axiom Optics 	\\
camera					&	(CMOS)			&					&				&				&  					&					& (OWL SW1.7HS)	\\
						&					&					&				&				&					&					&				\\
HiCIAO/					&      HgCdTe 			& 	$2048\times2048$ 	&    $18$			&   $15-30$		&  $<3$				&	$850-2500$		& Teledyne 		\\
CHARIS					&	(CMOS)			&					&				&				&  					&					& (HAWAII 2RG) 	\\
						&					&					&				&				&					&					&				\\
PyWFS		  			&      Si 				& 	$2560\times2160$ 	&    $6.5$			&   $1.2$			&  $100$				&   $350-1000$ 		&  Andor		   	\\ 
(previous)					&	(sCMOS)			&	($128\times128$ )	&				&				&  	($1700$)			& 					&	(Zyla)		\\
						&					&					&				&				&					&					&				\\
PyWFS	  				&      Si 				& 	$240\times240$ 	&    $24$			&   $<0.3$			&  $2000$				& $350-1000$			&  Firstlight Imaging	\\ 
(current)					&	(EMCCD)			&	($120\times120$)	&				& 				&  	($3700$)			& 					&  (OCAM$^{2}$K)	\\
						&					&					&				&				&					&					&				\\
Lucky 					&      Si 				&	$512\times512$ 	&    $16$			&   $<1$			&  $35$				& $300-1000$			&	 Andor		\\ 
imaging					&	(EMCCD)			&					&				&				&  					& 					&	(iXon3 897)	\\  
						&					&					&				&				&					&					&				\\
VAMPIRES				&      Si 				&	$512\times512$ 	&    $16$			&   $<1$			&  $56$				& $300-1000$			&	Andor		 \\ 
						&	(EMCCD)			&					&				& 				&  					& 					&  (iXon Ultra 897)	\\  
						&					&					&				&				&					&					&				\\
FIRST					&      Si 				&	$512\times512$ 	&    $16$			&   $<1$			&  $31.9$				& $300-1000$			&  Hamamatsu		 \\ 
						&	(EMCCD)			&					&				& 				&  					& 					&(ImagEM C9100-13)	\\ 
\enddata
\tablecomments{The values in the table are taken from manufacturer specifications/measurements. Values in parenthesis indicate the sub-framed (previous PyWFS) or binned (current PyWFS) image size and corresponding frame rate used. Read-out noise quoted for EMCCDs is with gain applied. Quantum efficiencies for these detectors are shown in Table~\ref{tab:QE}.}
\end{deluxetable*}

The light directed to the internal NIR camera passes through a filter wheel for spectral selectivity. The content of the filter wheel and the science light beamsplitter wheels are reported in Table~\ref{tab:splitters}. An image is formed on the internal NIR camera via a pair of IR achromatic lenses (see Section~\ref{sec:appendix} for details). The sampling and field-of-view on this and other cameras is summarized in Table~\ref{tab:sample}. The internal NIR camera is mounted on a translation stage allowing it to be conjugated to any location between the focal and pupil planes. 

The light reflected by the dichroic on the bottom bench (which splits the visible and IR channels) is directed towards a periscope which transports it to the upper bench. An achromatic lens ($50$~mm diameter, $f=500$~mm) is mounted in the periscope to reimage the pupil onto the top bench. A wheel hosts a range of dichroic beamsplitters at the focus of the beam on the top bench to select the spectral content to be directed towards the PyWFS (named WFS beamsplitter, content listed in Table~\ref{tab:splitters}). The light reflected is collimated by an achromatic lens ($f=200$~mm) and the pupil image is located on a piezo-tip/tilt mirror used to modulate the PyWFS. The reflected beam is routed to focusing optics which form a converging beam with $F/\#=40$ (a combination of achromatic lenses with focal lengths of $f=400$ and $f=125$~mm are used for this). A double pyramid prism shaped optic~\citep{esp10} is placed at the focus of the beam such that the vertex is on the optical axis. An image of the resulting four pupils is generated via one additional lens ($f=300$~mm) on the detector. The PyWFS is discussed in more detail in section~\ref{sec:WFS}.

\begin{deluxetable*}{ccccc}[ht!]
\tabletypesize{\footnotesize}
\centering
\tablecaption{Filter and beamsplitter wheel contents. \label{tab:splitters}}
\tablehead{
\colhead{Slot}		& 	\colhead{Internal camera filters	}		& 	\colhead{Science light beamsplitters}			&   \colhead{Wavefront sensor beamsplitter}	&  \colhead{VAMPIRES/FIRST splitter}}  \\ 
\startdata
1   				&  	Open							& 	AR-coated window ($900-2600$~nm)		&	Open						&    Silver mirror  					\\
2				&	T=$y$-band						&	$T, R=50\%$ ($900-2600$~nm)				&	Silver mirror					&    $T, R=50\%$, $600-950$~nm		\\
3 				&  	T=$J$-band						&   	$T=10\%$, $R=90\%$ ($900-2600$~nm)		&	$T, R=50\%$, $600-950$~nm		&    $T<700$~nm, $R>700$~nm		\\
4				&	T=$H$-band						&	$T=90\%$, $R=10\%$ ($900-2600$~nm)		&	$T<650$~nm, $R>650$~nm		&    $T>700$~nm, $R<700$~nm		\\
5				&     $50$~nm bandpass 					& 	$T>90\%$  ($1400-2600$~nm),  			&	$T<700$~nm, $R>700$~nm		&     AR-coated window 				\\
				&		at $1600$~nm							&     $R>95\%$ ($900-1400$~nm) 		&								&    	($600-950$~nm)				\\
6				&     	-								&      	$T>90\%$ ($1900-2600$~nm), 				& 	$T<750$~nm, $R>750$~nm 		&	Open						\\
				&									&      $R>95\%$ ($900-1900$~nm) 				&								&								\\  
7 				&      -								&	Gold mirror							&	$T<800$~nm, $R>800$~nm		&		-						\\
8 				&      -								&	-									&	$T<850$~nm, $R>850$~nm		&		-						\\
9  				&      -								&	-									 &	$T>750$~nm, $R<750$~nm		&		-						\\
10  				&      -								&	-									&	$T>800$~nm, $R<800$~nm		&		-						\\
11 				&      -								&	-									&	$T>850$~nm, $R<850$~nm		&		-						\\ \enddata
\tablecomments{The values in the table are based on the final measurements made by the manufacturer. T-Transmission, R-Reflection. AR-anti-reflection. Note item $5$ and $6$ of the science light beamsplitter have not been delivered yet.}
\end{deluxetable*}

\begin{deluxetable}{lccc}[h!]
\tabletypesize{\footnotesize}
\centering
\tablecaption{Plate scale and field-of-view. \label{tab:sample}}
\tablehead{
\colhead{Detector}				& 	\colhead{Sampling} 	&	\colhead{Field-of-view (")}        	\\  
							&	\colhead{(mas/pixel)}	&		}					\\
\startdata      
Internal NIR	  				& 	$12.1\pm0.1$		&	$4\times3$				\\
camera						&					&							\\
HiCIAO						&   	$8.3\pm0.1$		&	$>10\times10$				\\   
VAMPIRES (M)					&      -				& 	$0.08$--$0.46$  			\\
VAMPIRES (NM)				&	$6.0$				&	$2\times1$					\\
FIRST						&	-				&	$0.1$					\\
\enddata
\tablecomments{All values in this table were measured off-sky by moving the calibration source laterally in the input focal plane (AO$188$ focal plane) and determining this motion in pixels on the detector. This motion is converted to a plate scale based on the knowledge of the AO$188$ plate scale which is well known to be $1.865$~"/mm. This method yields consistent values to those obtained by looking at astrometric binaries. M-with aperture masks, NM-no masks. Range of field-of-view is dependent on choice of mask. Please see~\cite{norris15} for more details.}
\end{deluxetable}

The light which is not directed to the WFS is split between two modules: VAMPIRES and FIRST. The basic optical layouts are outlined here and details in regards to specifics, including calibration and performance are given in other publications~\citep{huby12,norris15}. For engineering purposes a grey beamsplitter is used to divide the light between VAMPIRES and FIRST ($50/50$), but can be swapped for several other optics if the science demands it (contents of the VAMPIRES/FIRST splitter wheel are shown in Table~\ref{tab:splitters}). 

The light transmitted by the splitter is used by the VAMPIRES module. It is first collimated (by BRTL2) and then passes through a series of optics including a liquid crystal variable retarder (LCVR), pupil masks, and spectral filters before being focused by a combination of a converging and diverging lens onto a low noise detector. A polarization beamsplitter, consisting of a polarizing beam cube and $3$ right-angled prisms is placed in the beam after the final focusing lens and is used to spatially separate the two orthogonally polarized components on the detector. The VAMPIRES module combines aperture masking interferometry with polarimetry. The sparse aperture masks are housed in the pupil wheel where an assortment of masks with various throughputs and Fourier coverage can be found (please see ~\cite{norris15} for details). The module operates on $50$~nm bandwidths of light selected within the $600$--$800$~nm range, via a set of spectral filters in order to maintain fringe visibility, while maximizing the signal-to-noise ratio. At $650$~nm VAMPIRES can achieve an angular resolution of $8.4$~mas (Fizeau criteria) and has a field-of-view in masking mode which ranges from $80$~mas to $460$~mas depending on the mask selected (larger fields-of-view, $\sim1$--$2"$ are possible in normal imaging mode, no mask). Rather than simply blocking the unwanted light from the pupil, the masks are reflective mirrors with transmissive sub-apertures so that the unwanted light is redirected to a pupil viewing camera (PtGrey, Flea, FL3-U3-13S2M-CS) which allows for fine alignment of the masks with the pupil. The pupil viewing mode of VAMPIRES is only used when aligning the masks. To utilize the light when pupil viewing is not being used a mirror is translated into the beam to direct the light to the Lucky imaging module/point spread function (PSF) viewer (described in detail in~\ref{sec:LFI}).

The high angular resolution imaging capability of VAMPIRES is boosted by an advanced polarimetric capability. This gains its strength from the multi-tiered differential calibration scheme which is utilized. Firstly, two quarter wave plates (HWP) mounted in front of the periscope on the bottom bench can be used to compensate for the birefringence induced by the mirrors in AO$188$ and SCExAO. The setting of these plates is done by careful calibration beforehand. Fast polarization modulation ($\sim50$~Hz) comes from the LCVR which is switched between every image. The analyzer splits the signal into two distinct interferograms on the detector with orthogonal polarizations. Finally, a HWP positioned in front of AO$188$, used for HiCIAO polarimetric imaging in the NIR, was determined to work well in the visible and is used as the main polarization switching component in VAMPIRES. The quarter wave plate (QWP) before the aperture mask wheel and the polarizer on the bottom bench are used for calibrating the polarization systematics with the internal calibration source and can be swung into the beam when off-sky (they are not used on-sky). For more details about the nested differential calibration procedure please refer to~\cite{norris15}.

The light reflected by the beam splitter is sent to the FIRST module. It is collimated before entering a polarizing beamsplitter cube. The main beam is reflected at $90^{\circ}$ onto a $37$ element segmented mirror (Iris AO~\citep{helm11}) which is conjugated to the pupil plane. A QWP placed between the polarizing beamsplitter and the segmented mirror is used to rotate the polarization so that the beam passes through the cube when reflected off the segmented mirror. A beam reducing telescope compresses the beam so that a single segment of the mirror has a one-to-one correspondance with a micro-lens in the micro-lens array (MLA) used for injecting into the bundle of single-mode polarization maintaining fibers (see Fig.~\ref{fig:scexao}). This architecture allows the segmented mirror to fine tune the coupling into each of the fibers with small tip/tilt control. Currently $18$ of the $30$ available fibers ($2$ sets of $9$ fibers each) are used and they transport the light to a separate recombination bench (see Fig.~\ref{fig:scexao}) where the interferograms are formed and data collected. A description of the recombination bench is beyond the scope of this work. The instrument offers an angular resolution of $9$~mas at $700$~nm with a $100$~mas field-of-view ($\sim6\lambda/D$ in radius). In addition, broadband operation from $600$--$850$~nm, with a spectral resolving power of $300$, offers a new avenue to maximizing data output while using standard bispectrum analysis techniques (and hence precision/contrast) while simultaneously allowing spectra to be collected.

FIRST has had several successful observing campaigns on close binary stars at Lick Observatory. It has now made the move from the Cassegrain focus of the Shane telescope to the gravitationally invariant Nasmyth platform of Subaru Telescope, which minimizes mechanical flexure and hence instrumental instabilities. A more comprehensive description of the instrument including the recombination bench, how it works and the initial science results is presented in~\cite{huby12, huby13}.

Despite all the advanced wavefront control, interferometers and coronagraphs in SCExAO, the performance of the system is highly dependent on the stability of the PSF. Vibrations can plague high-contrast imaging testbeds via flexure, windshake of mirrors, or moving parts (cryocoolers for example). For this reason several key efforts have been made to address PSF stability in SCExAO. Firstly four elastomer-based vibration isolators (Newport, M-ND$20$-A) are used to support the SCExAO bench at the Nasmyth platform. These elastomers have a natural frequency $9-12$~Hz and are used to damp high frequency mechanical vibrations from the surrounding environment. Secondly, the mounts for the OAP-based relay optics in the instrument were custom made from a single piece to minimize the drift of the PSF. Finally, the mounting scheme for the HiCIAO cryocooler was rebuilt such that the pump was connected to the dewer by springs and metal bellows only (soft connections with low spring constants for maximum damping). This reduced the tip/tilt jitter observed on the NIR internal camera by a factor of $10$ at $10$~Hz and up to $100$ at a resonant frequency of $23$~Hz~\citep{Jovanovic14a}. These steps have improved overall PSF stability against long term drifts and vibrations above a few Hz. However, there are small residual resonances which will be addressed with a linear quadratic Gaussian (LQG) controller~\citep{poy14} implemented to the LLOWFS loop in future.

\subsection{Internal calibration}
An important feature of any high-contrast testbed is the ability to internally calibrate it off-sky. The SCExAO instrument can be aligned/calibrated with its internal calibration source. The source can be seen in Fig.~\ref{fig:scexao} and consists of a standalone box which houses a super continuum source (Fianium - Whitelase micro) for broadband characterization, and two fiber coupled laser diodes ($675$~nm and $1550$~nm) for alignment. The light from the super continuum source is collimated and passes through a series of wheels which house neutral density filters for both visible and IR wavelengths as well as a selection of spectral filters. The light is coupled into an endlessly single-mode photonics crystal fiber (NKT photonics - areoGUIDE8) which transports the light to the SCExAO bench. The fiber is mounted on a translation stage and can be actuated into the focus of the AO$188$ beam (see Fig.~\ref{fig:scexao}) when internal calibration is performed or the instrument is not at the telescope. The endlessly single-mode fiber is ideal for this application as it offers a diffraction-limited point source at all operating wavelengths of the super continuum source and SCExAO ($600-2500$~nm). 

The effects of atmospheric turbulence can also be simulated. This is achieved by using the DM to create a large phase-screen with the appropriate statistical fluctuations. A single-phase screen Kolmogorov profile is used where the low spatial frequencies can be attenuated, to mimic the effect of an upstream AO system like AO$188$. A $50\times50$ pixel sub-array (corresponding to the actuators of the DM) is extracted from the larger phase screen. By scanning the sub-array across the phase map, a continuous and infinite sequence of phase screens can be generated. The amplitude of the RMS wavefront map, the magnitude of the low spatial frequency modes and the speed of the sub-array passing over the map (i.e. windspeed) are all free parameters that can be adjusted. This simulator is convenient as it allows great flexibility when characterizing SCExAO modules. Note that due to the limited stroke of the DM (which is discussed in section~\ref{sec:DM}), the turbulence simulator cannot be used to simulate full seeing conditions but provides a level of wavefront perturbation that is representative of post AO$188$ observing conditions. Finally, although the simulator can provide turbulence with the correct spatial structure pre or post AO, it does not take into account the temporal aspects of the correction provided by an upstream AO system (i.e. AO corrects low temporal frequencies leaving only higher frequencies).  

\subsection{Instrument throughput}
To plan observing schedules and determine the limitations of the instrument, the throughput was accurately characterized. As SCExAO has many branches, Tables \ref{tab:throughputIR} and \ref{tab:throughputVIS} summarize the measured internal system throughput for each. Although not explicitly listed, the throughputs include all flat mirrors required to get the light to a given module. In addition Table~\ref{tab:throughputTel} highlights the measured throughputs of the optics upstream from SCExAO which includes the effects of the atmosphere, telescope, AO$188$ HWP, atmospheric dispersion compensator (ADC), AO$188$ optics as well as HiCIAO. To determine the total throughput to a given detector plane of the instrument, one should first use Table~\ref{tab:splitters} to select the appropriate beamsplitter, and then find the corresponding throughput for that branch from Table~\ref{tab:throughputIR} or \ref{tab:throughputVIS}. This value should then be multiplied by the relevant throughputs in Table~\ref{tab:throughputTel}. To convert to photoelectrons, the quantum efficiencies for the detectors used in SCExAO are listed in Table~\ref{tab:QE}. Finally, the throughputs of the focal plane masks used for the coronagraphs are not shown in Table~\ref{tab:throughputIR} as they are specifically designed to attenuate the light. Indeed, the throughput of these masks depends on the distance from the optic axis and for this we refer the reader to literature such as~\cite{guyon05}.

\begin{deluxetable}{lcc}
\tabletypesize{\footnotesize}
\centering
\tablecaption{Throughput of the various arms of the IR channel \\of SCExAO measured in the laboratory from \\the internal calibration source. \label{tab:throughputIR}}
\tablehead{
\colhead{Path (elements)}	& 	\colhead{Band}		& 	\colhead{Throughput (\%)}} 		\\   

\startdata
\bf{From calibration}   		&  					& 					\\
\bf{source input to....}		&					&					\\
						&					&					\\
\bf{Internal NIR camera}	 	&  					& 					\\
(OAPs, DM, SLB2)			&  	$y$				&   	35.3				\\
(OAPs, DM, SLB3)			&	$y$				&	60.6				\\
(OAPs, DM, SLB4)			&      $y$				& 	4.5				\\
(OAPs, DM, SLB7)			&      $y$				& 	71.1				\\
(OAPs, DM, SLB2)			&     	$J$				&   	27.2				\\  
(OAPs, DM, SLB3)			&     	$J$				&   	53.7				\\  
(OAPs, DM, SLB4)			&     	$J$				&   	5.9				\\  
(OAPs, DM, SLB7)			&     	$J$				&   	50.4				\\   
(OAPs, DM, SLB2)			&     	$H$				&   	32.8				\\  
(OAPs, DM, SLB3)			&     	$H$				&   	56.2				\\  
(OAPs, DM, SLB4)			&     	$H$				&   	9.4				\\  
(OAPs, DM, SLB7)			&     	$H$				&   	63.4				\\  
						&					&					\\
\bf{Facility science instruments}	&					&					\\
(OAPs, DM, SLB1)			&	$y$				&	72.1				\\
(OAPs, DM, SLB2)			&  	$y$				&   	34.9				\\
(OAPs, DM, SLB3)			&	$y$				&	7.6				\\
(OAPs, DM, SLB4)			&      $y$				& 	67.2				\\
(OAPs, DM, SLB1)			&     	$J$				&   	66.9				\\  
(OAPs, DM, SLB2)			&     	$J$				&   	33.4				\\  
(OAPs, DM, SLB3)			&     	$J$				&   	6.2				\\  
(OAPs, DM, SLB4)			&     	$J$				&   	60.5				\\  
(OAPs, DM, SLB1)			&     	$H$				&   	65.2				\\  
(OAPs, DM, SLB2)			&     	$H$				&   	31.8				\\  
(OAPs, DM, SLB3)			&     	$H$				&   	5.0				\\  
(OAPs, DM, SLB4)			&     	$H$				&   	58.2				\\  
						&					&					\\

\bf{Single-mode injection}		&					&					\\
(OAPs, DM, FIB)			&  	$J$				&	77.0				\\
(OAPs, DM, FIB)			&  	$H$				&	78.3				\\
						&					&					\\

\bf{Components in isolation} 	&					&					\\
(PIAA+binary Mask+IPIAA)	&	$y$				&	53.2				\\
(PIAA+binary Mask+IPIAA)	&	$J$				&	52.2				\\
(PIAA+binary Mask+IPIAA)	&	$H$				&	52.5				\\	
\enddata
\tablecomments{OAPs-refers to off-axis parabolic mirrors used to relay the light, DM-The window and mirror of the DM, SLB-science light beamsplitter, the number designates the slot (see Table~\ref{tab:splitters}), FIB-Fiber injection beamsplitter.}
\end{deluxetable}

\begin{deluxetable}{lcc}
\tabletypesize{\footnotesize}
\centering
\tablecaption{Throughput of the various arms of the visible \\channel of SCExAO measured in the laboratory \\ from the internal calibration source. \label{tab:throughputVIS}}
\tablehead{
\colhead{Path (elements)}				& 	\colhead{Band}		& 	\colhead{Throughput (\%)}}	\\  
\startdata
\bf{From calibration}   				&  					& 				\\
\bf{source input to....}				&					&				\\
								&					&				\\
\bf{Pre-WFSBS}					&					&				\\
(PTTB)							&	$R$				&	57.8			\\
								&	$I$				&	65.4			\\	
								&	$z$				&	65.6			\\
								&					&				\\

\bf{PyWFS camera} 					&					&				\\
(PTTB + PyWFS optics 				&	$R$				&	43.0			\\
WFSBS 2)						&	$I$				&	47.1			\\
				          			&	$z$				&	47.6			\\ 
(PTTB + PyWFS optics,				&	$R$				&	20.5			\\
 WFSBS 3)						&	$I$				&	23.0			\\
		         					&	$z$				&	25.2			\\ 
(PTTB + PyWFS optics, 				&	$R$				&	42.4			\\
WFSBS 4, 9, 10 or 11)  				&					&				\\
(PTTB + PyWFS optics,				& 	$I$				&	49.1			\\
 WFSBS 4, 5, 6 or 7) 				&					&				\\    			
(PTTB + PyWFS optics, 				&	$z$				&	48.7			\\
WFSBS 4, 5, 6, 7 or 8) 				&					&				\\
       								&					& 				\\

\bf{VAMPIRES camera}				&					&  				\\
(PTV + WFSBS 3)					&	$R$				&	12.6			\\
								&	$I$				&	15.3			\\
 								&	$z$				&	15.2			\\
(PTV + WFSBS 5, 6, 7 or 8) 			&	$R$				&	25.3			\\
(PTV + WFSBS 8) 					&	$I$				&	27.5			\\	
(PTV + WFSBS 9, 10 or 11) 			&	$z$				&	30.6)			\\
								&					&				\\
\bf{FIRST input}					&					&				\\
(PTF + WFSBS 3)					&	$R$				&	12.6			\\
	 							&	$I$				&	15.3			\\
								&	$z$				&  	13.4			\\
(PTF + WFSBS 5, 6, 7 or 8)			&	$R$				&	27.1			\\
(PTF	 + WFSBS 8) 					&	$I$				&	29.0			\\
(PTF + WFSBS 9, 10 or 11)			&	$z$				&	28.6			\\
								&					&  				\\
\enddata
\tablecomments{OAP1-refers to the first off-axis parabolic mirror used to relay the light, DM-The window and mirror of the DM, WFSBS- Wavefront sensor beamsplitter, the number designates the slot (see Table~\ref{tab:splitters}), BRTL1-beam reducing telescope lens $1$ which is located in the periscope. Path to top bench (PTTB): OAP1, DM, dichroic, QWP $\times 2$, periscope mirrors, BRTL1. Path to VAMPIRES (PTV): OAP1, DM, dichroic, QWP $\times 2$, periscope mirrors, BRTL1 \& 2, 50/50 BS and focusing lenses. Path to FIRST (PTF): OAP1, DM, dichroic, QWP $\times 2$, periscope mirrors, BRTL1, 50/50 BS and collimating lens. Note the throughput is only quoted to the input of FIRST. Also, the throughput of the aperture masks for VAMPIRES were not included, please see~\cite{norris15} for these values.}
\end{deluxetable}

\begin{deluxetable}{lcccccc}[ht!]
\tabletypesize{\footnotesize}
\centering
\tablecaption{Throughput of the atmosphere, telescope, \\AO$188$ HWP,  atmospheric dispersion compensator (ADC), \\AO$188$ optics and HiCIAO as a percentage. \label{tab:throughputTel}}
\tablehead{
\colhead{Band}		& 	\colhead{Atmos.}		& 	\colhead{Tele.}			& 	\colhead{HWP}				& 	\colhead{ADC}		&	\colhead{AO$188$}	&	\colhead{HiCIAO}}	\\  
				&						& 						&							&					&					&			\\  
$R$				&	91					&	82					&		93					&	95				&	68				&	-		\\
$I$				&	96					&	75					&		95					&    	95				& 	69				&	-		\\	
$z$				&	97					&	79					&		95					& 	95				&	69				&	-		\\
$y$				&	98					&	80					&		94					&	95				&	72				&	-		\\
$J$				&	95					&	91					&		96					&	94				& 	77				&	87		\\
$H$				&	97					&	92					&		94					&	92				&	79				&	78		\\
$K$				&	92					&	93					&		92					&	70				&	82				&	82		\\
\enddata
\tablecomments{The atmospheric transmission in R, I, Z bands is estimated from \cite{buton12}. The throughput of the atmosphere in the y, J, H and K bands assumes an airmass of 1.5 and precipitable water vapor content of 1.6 mm (data taken from Gemini Observatory website)~\citep{lord92}. The throughput of the telescope is calculated from the reflectivity of the material used for the coating and does not include effects like diattenuation from M$3$ for example. The throughput for the HWP and ADC were measured on-sky. The throughputs for R, I, Z, and y band for AO$188$ have been calculated from materials. J, H and K band throughputs for AO$188$ were measured.}
\end{deluxetable}

\begin{deluxetable}{lccccccc}[ht!]
\tabletypesize{\footnotesize}
\centering
\tablecaption{Quantum efficiency as a function of waveband \\for the detectors in SCExAO as a percentage. \label{tab:QE}}
\tablehead{
\colhead{Detector}		&	$R$		& 	$I$		&		$z$		&	$y$		&	$J$		&	$H$		&     $K$ 	}	\\  
					&			&			& 				&			&			&			&			\\  
Internal NIR camera		&	-		&	-		&		-		&	58		&	73		&	73		&	-		\\
LOWFS camera		&	-		&	-		&		-		&    	58		& 	73		&	73		&	-		\\	
HiCIAO/CHARIS		&	-		&	-		&		-		& 	-		&	46		&	68		&	68		\\
PyWFS (current)		&	98		&	96		&		77		&	-		&	-		&	-		&	-		\\
Lucky imaging			&	93		&	74		&		47		&	-		& 	-		&	-		&	-		\\
VAMPIRES			&	93		&	74		&		42		&	-		&	-		&	-		&	-		\\
FIRST				&	91		&	70		&		42		&	-		&	-		&	-		&	-		\\
\enddata
\tablecomments{Values were taken from manufacturer specifications.}
\end{deluxetable}

From the tables we can determine the throughput to several focal planes within SCExAO as an example of the system performance and a demonstration of how to use the tables. One commonly used observing mode involves directing the majority of the post-coronagraphic light towards HiCIAO for H-band imaging. In this case one would select the $90\%$ transmitting beamsplitter in the science light beamsplitter wheel (Slot $4$ from Table~\ref{tab:splitters}). In this case the throughput of SCExAO would be $9.4\%$ to the internal NIR camera and $58.2\%$ to HiCIAO (from Table~\ref{tab:throughputIR}). To determine the total throughput from the top of the atmosphere, one would need to multiply the SCExAO throughputs by those listed for the atmosphere ($97\%$), telescope ($92\%$), HWP ($94\%$), ADC ($92\%$), AO$188$ ($79\%$) and HiCIAO optics/filters ($78\%$) by those listed in Table~\ref{tab:throughputTel}, which yields values of $4.5\%$ and $27.7\%$ respectively. From here one would need to take into account the losses of the focal plane and Lyot masks which are coronagraph specific. Finally, it is possible to use the quantum efficiencies listed in Table~\ref{tab:QE} to convert the throughput of the instrument into the number of photoelectrons on the HiCIAO detector for a given magnitude target.  

In addition to flux at the detector plane, the throughput to the primary wavefront sensor is a useful piece of information for appropriate target selection. Since the PyWFS does most of its sensing at $850$--$900$~nm, one could operate with an $800$~nm short pass filter in the wavefront sensor beamsplitter wheel (slot 7 from Table~\ref{tab:splitters}). The throughput of SCExAO to the WFS detection plane with this splitter would be $48.7\%$ in the z-band (from Table~\ref{tab:throughputVIS}). The associated throughput as measured from the top of the atmosphere would be $23.2\%$ (including the z-band throughputs from Table~\ref{tab:throughputTel}). This relatively low throughput is attributed to the fact the PyWFS is a long way downstream of the telescope and even within SCExAO. In the future, if SCExAO were to be rebuilt, one could imagine moving the WFS to be the first element after the telescope and ADC which would no doubt improve the throughput by a factor of $2$ at least. This would allow the PyWFS to operate with high performance on a $1$ magnitude fainter star. 

\section{The functionalities of SCExAO}\label{sec:fund}
Key to the successful implementation of a high performance coronagraph or interferometer for high-contrast imaging is a wavefront control system to correct for both atmospheric as well as instrumental aberrations. Wavefront control comprises two primary components: sensing and correction. The former is taken care of by a wavefront sensor, a device designed to map the spatial profile of the phase corrugations while the latter by an adaptive element such as a DM. In this section we describe how the elements of SCExAO function. 

\subsection{Wavefront control}
\subsubsection{Wavefront correction}\label{sec:DM}
\begin{figure}[t!]
\centering 
\includegraphics[width=0.90\linewidth]{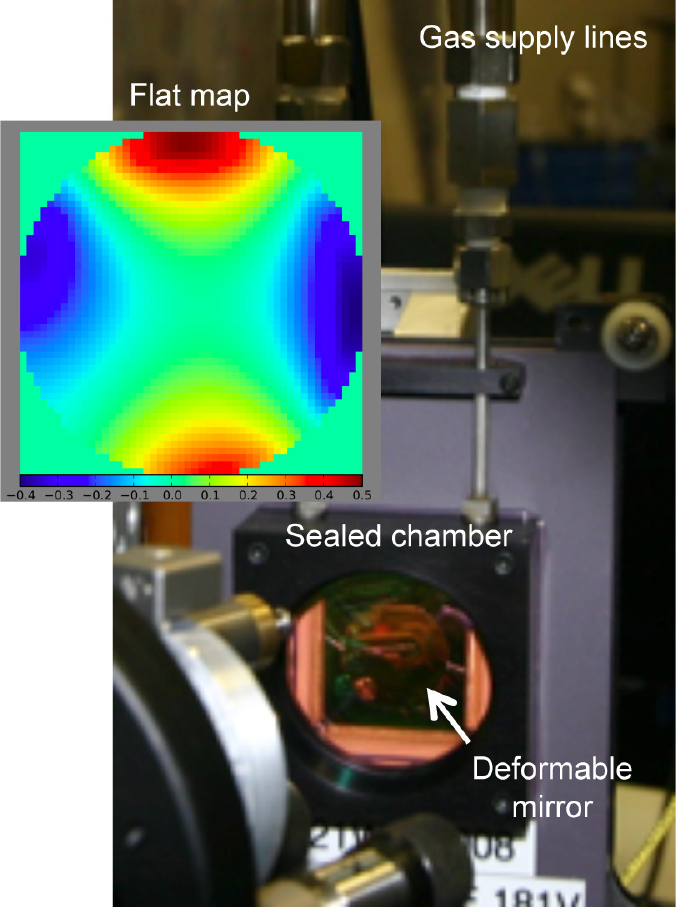}
\caption{\footnotesize $2$k DM mounted in the SCExAO instrument. The gas supply lines to the sealed chamber can be seen along with the gold surface of the DM itself. Inset: The map applied to the DM in order to compensate for the DM surface figure. The map shows the magnitude of the modulation of the actuators in microns. The window is $50$ actuators in diameter corresponding to the functioning region of the DM. The map required to flatten the DM surface is dominated by the Zernike which represents astigmatism.}
\label{fig:DM}
\end{figure}

The deformable mirror is the proverbial heart  of any adaptive optics system. The $2$k DM used in SCExAO (Figure~\ref{fig:DM}) is manufactured by Boston Micromachines Corporation based on MEMS technology.
The DM is enclosed in a sealed chamber in order to control its environment. The critical parameter to control is the humidity level and is kept below $20\%$ as advanced aging of MEMS actuators has been observed when high voltages ($180$~V is the maximum that can be applied), required to actuate individual elements, are applied in a moist environment~\citep{shea04}. A desiccant is used to filter the compressed air of moisture. A circuit monitors both humidity and pressure and is setup to interlock the power to the DM based on these two metrics (see Section~\ref{sec:appendix}). 

\begin{figure*}[ht!]
\centering 
\includegraphics[width=0.80\linewidth]{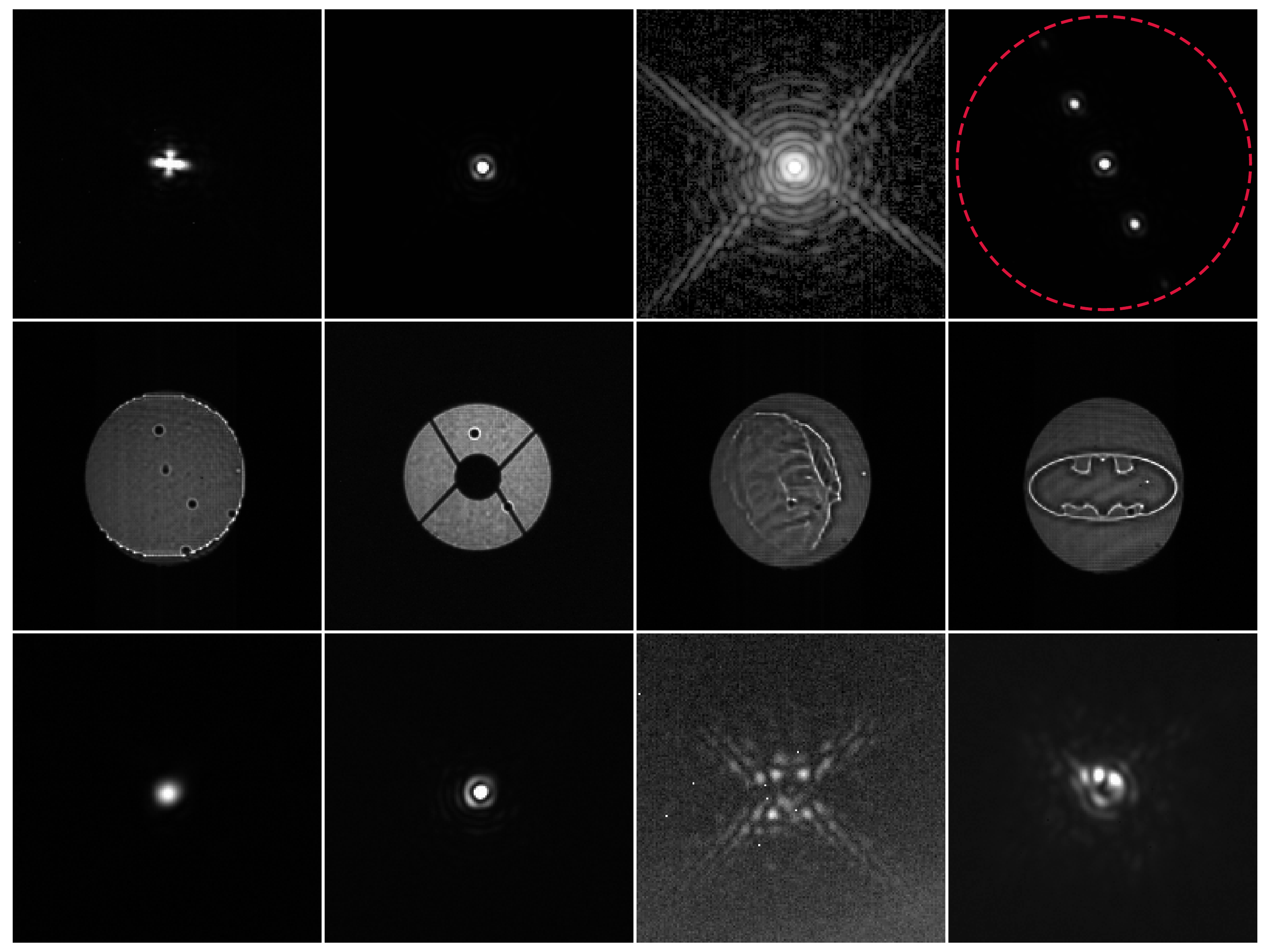}
\caption{\footnotesize Top row: PSF of SCExAO at $1550$~nm with the unpowered DM surface (left). Strong astigmatism is clearly visible. An image of the instrument PSF taken with the optimum flat map applied with linear (middle-left) and logarithmic (middle-right) scaling. The image is diffraction limited and numerous Airy rings can be seen. Image with several artificial speckles applied (right). The dashed ring designates the edge of the control region of the DM corresponding to a radius of $22.5~\lambda/D$ or $900$~mas at $1550$~nm. Second row: Pupil images showing the unmasked surface of the DM with 5 dead actuators (left), the internal spider mask in place masking several actuators (middle-left), an image of the PI Guyon (middle-right) and the bat symbol (right). These two images demonstrate the resolution of the DM. Bottom row: Image of the PSF taken in the laboratory with a laser at $1550$~nm, after the conventional PIAA lenses (left), with the IPIAA lenses (middle-left) as well as with the achromatic focal plane mask (AFPM) ($1.9~\lambda/D$ IWA) (middle-right). (right) An image with the PIAA, IPIAA and AFPM is shown taken on-sky on the night of the $25^{th}$ of July $2013$ with the full H-band.}
\label{fig:PSF}
\end{figure*}

The chamber window is optimized for transmission across the entire operating range of SCExAO, $550-2500$~nm (anti-reflection coated, IR fused silica). The DM's silicon membrane is gold coated and hosts $2000$ actuators within the $18$~mm diameter beam (there are others outside the pupil but they are not connected). The membrane is bonded to the $\sim250~\mu$m square-shaped actuators which are on a pitch of $400~\mu$m. This means that there are $45$ and $45.5$ actuators across the $18$~mm beam in the vertical and horizontal directions respectively. The number of actuators across the DM defines the highest spatial frequency components which can be probed/controlled by the DM and hence the region of control in the focal plane. For our case the $45$ actuator pupil diameter means that the fastest modulation that can be Nyquist sampled by the DM consists of $22.5$ cycles across the diameter. This means that spatial frequencies out to $22.5~\lambda/D$ from the PSF can be addressed, which defines the radius of the control region in the focal plane ($0.9"$ in H-band, highlighted in Fig.~\ref{fig:PSF}). Figure~\ref{fig:PSF} shows the surface of the DM, where the $5$ dead actuators (actuators that cannot be modulated) are clearly visible. Fortunately, it was possible to position three of the actuators behind the secondary or outside the pupil and one is partially blocked by the spiders of the telescope, leaving only one dead segment in the illuminated pupil. Dead segments compromise the maximum contrast achievable post coronagraph as they diffract light and hence the $1$--$2$ illuminated for SCExAO is close to ideal. Note, most coronagraph Lyot stops currently installed in SCExAO do not mask out the actuators, but there are plans to include such masks in the future. In addition the resolution of the DM can be qualitatively examined in Fig.~\ref{fig:PSF} where a bit map image of the PI, Guyon's face and the bat symbol have been imprinted in phase. The camera used to capture the image is not conjugated to the plane of the DM so that phase information gets recorded as amplitude information on the camera. The resolution of the images generated by the DM demonstrates the high level of sophistication MEMS technology has reached.

The surface of the unpowered DM is not flat. The inset to Figure~\ref{fig:DM} shows the voltage map that needs to be applied to obtain a flat DM in units of microns of stroke, known as a flatmap. The distortion of the surface of the DM is clearly dominated by astigmatism. Images of the PSF of SCExAO taken without and with the flat map applied are shown in the top left panels of Fig.~\ref{fig:PSF}. The image without the flatmap is consistent with the presence of strong astigmatism. The PSF post flatmap is diffraction-limited. It is clear that the maximum stroke required for correction near the edges approaches $0.5~\mu$m. This means that $25\%$ of the $2~\mu$m stroke of the actuators is used up to flatten the DM. 

\subsubsection{Wavefront sensing}\label{sec:WFS}
\begin{figure}[t!]
\centering 
\includegraphics[width=0.95\linewidth]{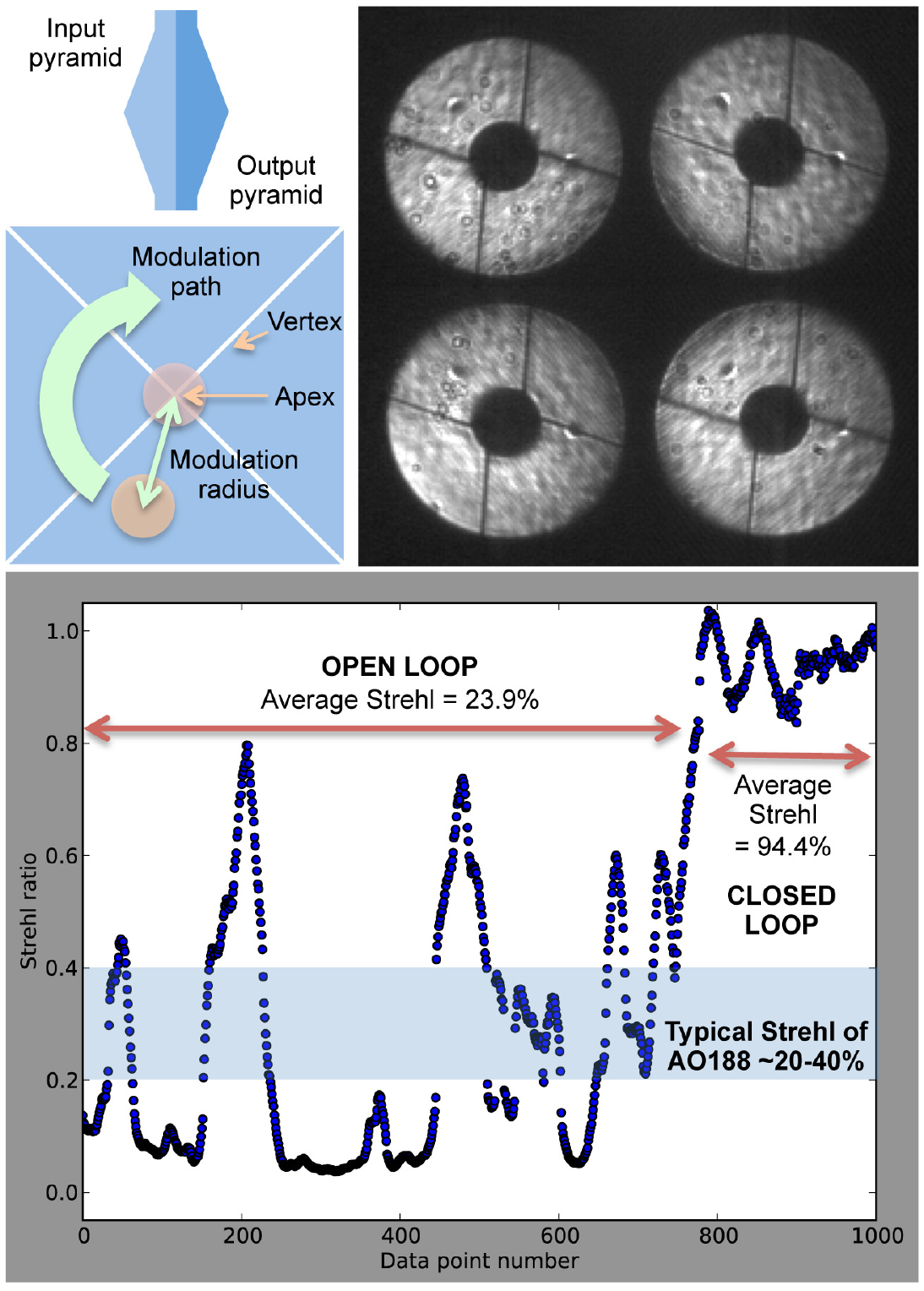}
\caption{\footnotesize (Top left) A side view of the double pyramid. A very shallow angle ($\sim1^{\circ}$) is required which is hard to produce so two steeper pyramids are used such that the cumulative effect of refraction between them is equivalent to a single shallower sloped pyramid optic~\citep{esp10}. Below this a front view of the pyramid is shown. The circles represent the PSF position without modulation (red spot) and with modulation (orange spot). The green arrow shows the path of the PSF across the front face of the pyramid optic when modulation is applied. (Top right) An image of the four pupils generated by the pyramid optic taken with modulation applied. (Bottom image) Strehl measured for the internal source with $300$~nm RMS wavefront error applied to the turbulence simulator and the PyWFS loop open and then closed. NB, the Strehl ratio calculation procedure has a limited accuracy of about $5-7\%$ when the Strehl ratio is $>90\%$. Hence values $>1$ are possible in this regime as shown in the above chart and should simply be interpreted as regions of high Strehl, with no emphasis put on the exact value.}
\label{fig:PyWFS}
\end{figure}

Wavefront correction within the SCExAO instrument comes in two stages: low spatial and temporal frequencies are partially corrected by AO$188$ prior to being injected into SCExAO where a final correction including higher order modes is implemented. In good seeing conditions AO$188$ can offer $30-40\%$ Strehl ratios in the H-band. The high-order wavefront correction, which is the focus of this section is facilitated by a pyramid wavefront sensor (PyWFS). The PyWFS is chosen because of its large dynamic range and its sensitivity properties~\citep{guyon05}. In its standard implementation, a tip/tilt mirror modulates the location of the PSF in a circular trajectory which is centered on the apex of the pyramid. This implementation has been used to correct seeing-limited light to Strehl ratios as high as $90\%$ in very good seeing on LBTAO~\citep{esp11} and MagAO~\citep{close13}. By removing modulation however, the range over which the sensor responds linearly to aberrations is greatly reduced. Hence to utilize a PyWFS without modulation the wavefronts must be partially corrected by another sensor initially. With AO$188$ providing such a correction upstream, this implementation is possible with the PyWFS on SCExAO. SCExAO incorporates a piezo-driven mirror mount (shown in Fig.~\ref{fig:scexao}) to provide the modulation functionality, with the driver carefully synchronized to the frame rate of the camera. The implementation allows for continuous changes of the modulation radius (see Fig.~\ref{fig:PyWFS}). Such an architecture enables the possibility to start with a modulated PyWFS that has a larger range of linearity and slowly transition to a non-modulated sensor for maximum sensitivity and the highest Strehl ratio, as wavefront errors are gradually reduced.

The PyWFS has undergone laboratory and initial testing on-sky. In its initial format it exploited a micro-lens array instead of a pyramid optic as it was easier to obtain (pyramid optics need to be custom made). However, it was determined that although small micro-lenses have good inter-lens quality, which keeps diffraction effects at a minimum, they have a limited field-of-view which limits their use with modulation. On the other hand, larger micro-lenses remove this limitation but the inter-lens quality is poor and results in strong diffraction. Hence it was not possible from the $2$ micro-lens arrays tested to simultaneously obtain a large field-of-view and low diffraction (there may indeed be micro-lenses on the market that can achieve both)~\citep{clergeon14}. For this reason a dedicated pyramid optic was obtained. The pyramid optic presented here is a double pyramid, as shown in Figure~\ref{fig:PyWFS} and is a replica of the one used on MagAO~\citep{esp10,close13}. The pyramid optic segments the PSF and generates $4$ images of the pupil on the camera (see Fig.~\ref{fig:PyWFS}). The OCAM$^{2}$k from FirstLight imaging (see full specs. in Table~\ref{tab:detectors}), is used as the detector and is capable of photon counting, with low read noise, high frame rate, and low latency ($<1$ frame) which enables the correction of high temporal frequencies and allows operation on faint guide stars. To facilitate full speed PyWFS loop operation (limited by camera frame rate), fast computations are required and for this a bank of GPUs is utilized. Details of the control loop architecture for the PyWFS are beyond the scope of this work. Here we focus on the recent performance of the loop. 

In laboratory testing, the PyWFS successfully closed the loop with modulation (amplitude of $1.7~\lambda/D$) on up to $830$ modes. Our modal basis consisted of linearly-independent modes obtained by singular value decomposition of an input basis consisting of $5$ Zernike coefficients (tip, tilt, focus, astigmatisms) and the remainder Fourier modes (sine waves) up to a given spatial frequency of the DM. Non-orthogonality between input modes was addressed by rejecting (not controlling) low-eigenvalue modes. This basis set was chosen purely for convenience. This was achieved with the turbulence simulator generating $300$~nm RMS wavefront error maps with a wind speed of $5$~m/s and the low order spatial frequencies scaled to $30\%$ of the Kolmogorov power spectrum value to simulate upstream low order wavefront correction. Light with wavelengths between $800$ and $940$~nm was used for the tests and the loop was run at $1$~kHz. The Strehl of the images captured by the internal NIR camera was measured in the open and closed-loop regime and the results depicted in Fig.~\ref{fig:PyWFS}. The average Strehl in the open-loop regime was $23\%$ which is consistent with the value predicted from Marechal's approximation~\citep{hardy98} ($22.7\%$) given $300$~nm of wavefront error at $1550$~nm. When the loop is closed on $830$ modes, the Strehl improved to $>90\%$ on average confirming that given realistic post-AO correction, the PyWFS can indeed achieve extreme AO performance as required. 

The non-modulated mode was also tested with the turbulence simulator in the laboratory. However, due to the smaller linear range of the sensor without modulation the amplitude of wavefront errors was reduced. For the purposes of the test it was set to $60$~nm (windspeed $5$~m/s and low order Kolmogorov frequencies set to $30\%$). The loop was successfully closed at $1.5$~kHz on the turbulence simulator on $1030$ modes (which include $5$ Zernikes). In this regime, it was clear that the speckles in the halo surrounding the PSF were held very static. The loop speed is constantly being improved by optimizing the code and removing delays. An operational speed as high as $3.5$~kHz with minimal delay should be possible in the near future. Since the WFS detector operates in photon-counting mode, the loop will be able to run at full speed ($3.5$~kHz) even on faint targets (I-mag $\sim10$) albeit at lower loop gain. Fainter targets still will require lower loop speeds. An optimal modal gain integrator will also be implemented soon. 

Thus far the PyWFS has undergone some initial on-sky testing and has performed well on up to $130$ modes (with modulation, 1.7 $\lambda/D$ radius). However as the speed of the loop was the same as AO$188$ (i.e. $1$~kHz) and there were less modes than are corrected by AO$188$ a negligible improvement in Strehl was observed. Most of the gain came in the form of reduced tip/tilt jitter which was clearly visible in long integration time images. With further improvements in the AO loop code, the PyWFS should perform as demonstrated in the laboratory more recently, on-sky.

Non-common path and chromatic low-order errors between the visible PyWFS and the IR coronagraphs, are measured with the LLOWFS on the IR channel. The LLOWFS utilizes the light diffracted by the focal plane masks of the coronagraphs (discussed in detail in sections~\ref{sec:cor}), which is otherwise thrown away. A reflective Lyot stop is used to direct the diffracted light to the LOWFS camera~\citep{singh14}. In this way a reimaged PSF formed on the camera can be used to drive low-order, including tip/tilt, corrections by looking at the presence of asymmetries in the image. It has been tested thoroughly both in the laboratory and on-sky. Indeed it was used on-sky in conjunction with the vector vortex coronagraph on Vega, on the nights of the $14/15^{th}$ April, $2014$, and produced residual RMS tip/tilt wavefront errors of $0.01\lambda/D$~\citep{singh15}. 

In addition to the PyWFS and LLOWFS we are testing other wavefront sensing techniques. One such technique is known as focal plane wavefront sensing which exploits eigenphase imaging techniques~\citep{martinache13}. The focal plane wavefront sensor relies on establishing a relationship between the phase of the wavefront in the pupil plane and the phase in the Fourier plane of the image. Although it has a limited range of linearity ($\sim\pi$~radians), which means that the wavefront must first be corrected to the $~40\%$ Strehl ratio level before this sensor can be utilized, it can boost the Strehl ratio to $>95\%$ in the H-band by correcting low order modes. In addition, it operates just as effectively in the photon noise regime and is extremely powerful as non-common path errors are eliminated. This wavefront sensor is currently under development and has been successfully tested on both the internal calibration source, in which case the aberrations due to the internal SCExAO optics were corrected as well as on-sky, where the static aberrations due to the telescope, AO$188$ optics and SCExAO were all corrected. Some additional detail of this work can be found in \cite{martinache14a}.

\subsubsection{Coherent speckle modulation and control}\label{sec:CDI}
As a booster stage to the primary wavefront control loops, SCExAO makes use of coherent speckle modulation and control to both measure and attenuate residual starlight in the instrument's post-coronagraph focal plane. The $2$k DM actuators are used to remove starlight from a pre-defined region, referred to as the dark hole~\citep{malbet95}. Active modulation, induced by the $2$k DM, creates coherent interferences between residual speckles of unknown complex amplitude and light added by modifying the DM's shape (this component's complex amplitude is known from a model of the DM response and the coronagraph optics). By iterating cycles of measurement and correction, starlight speckles that are sufficiently slow to last multiple cycles are removed from the dark hole area. This approach, developed and perfected in the last $20$~yrs~\citep{malbet95, borde06, giveon07, guyon10b, codona13}, is well suited to high-contrast imaging as it effectively targets slow speckles, which are the dominant source of confusion with exoplanets. It also allows coherent differential imaging (CDI), a powerful post-processing diagnostic allowing true sources (incoherent with the central starlight) to be separated from residual starlight~\citep{guyon10b}. Compared to passive calibration techniques, such as angular differential imaging, CDI offers more flexibility, and achieves faster averaging of speckle noise. This is especially relevant at small angular separations, where ADI would require very long observation time to achieve the required speckle diversity. An example of a pair of speckles being generated by a periodic corrugation applied to the DM and used for starlight suppression is shown in the top right inset of Fig.~\ref{fig:PSF}. 

\begin{figure}[t!]
\centering 
\includegraphics[width=0.80\linewidth]{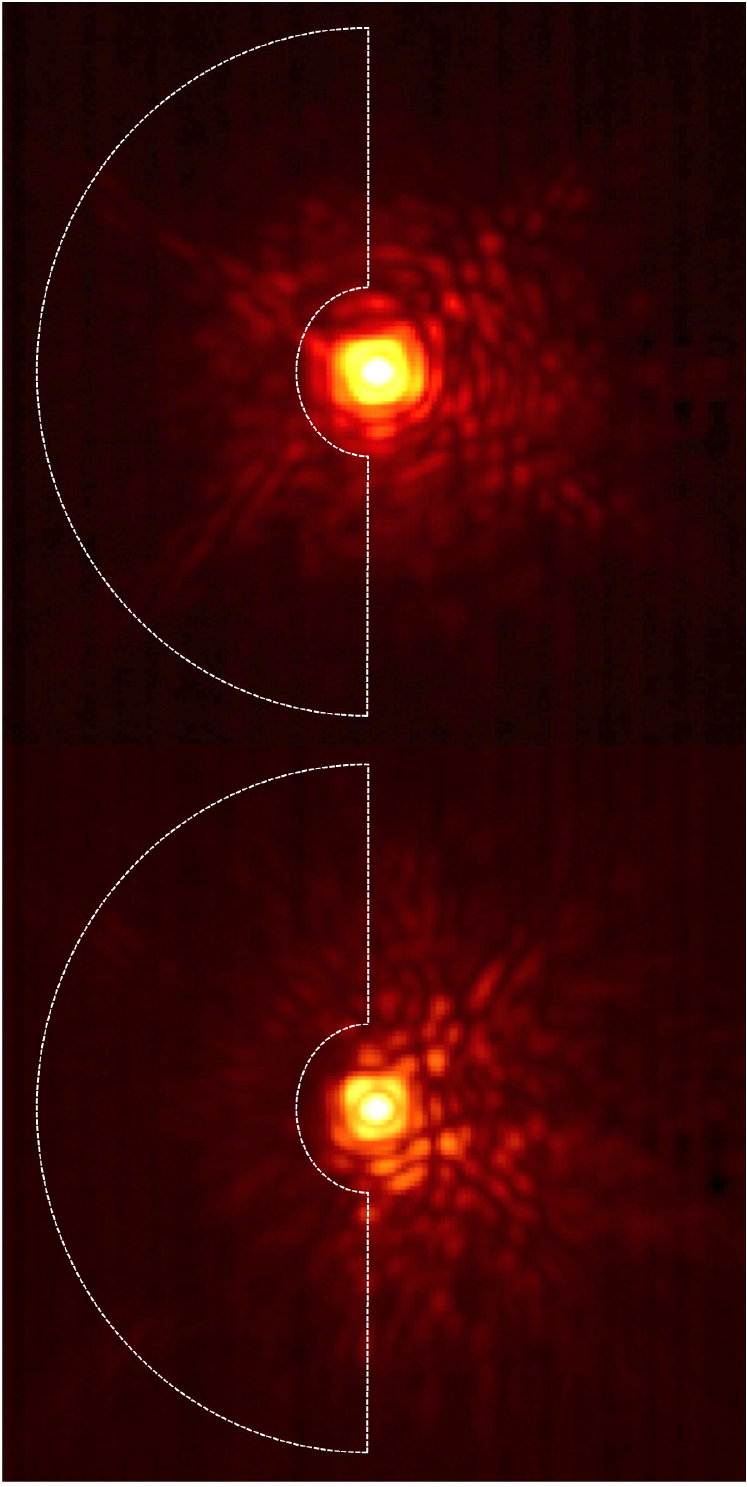}
\caption{\footnotesize Top: RX Boo with no speckle nulling applied. Bottom: RX Boo with speckle nulling performed on the region enclosed by the white dashed line. Each image is a composite of $5000$, $50~\mu$s frames which have been shifted and added together. Each panel has a square root stretch applied to it and the maximum and minimum values are clipped for optimum viewing contrast.}
\label{fig:sn}
\end{figure}

In SCExAO, coherent speckle control is implemented as discrete speckle nulling: the brightest speckles are identified in the dark hole region, and simultaneously modulated by the $2$k DM, revealing their complex amplitudes. The $2$k DM nominal shape is then updated to remove these speckles, and successive iterations of this loop gradually remove slow and static speckles. While discrete speckle nulling is not as efficient as more optimal global electric field inversion algorithms, it is far easier to implement and tune and thus more robust for ground-based systems which have much larger wavefront errors than laboratory testbeds or space systems. This approach has been validated on SCExAO both in the laboratory~\citep{martinache12} and on-sky~\citep{martinache14b} and is a means of carving out a dark hole on one side of the PSF to boost contrast in that region. A recently taken image demonstrating the successful implementation of speckle nulling without a coronagraph on RX Boo is shown in the lower panel of Fig.~\ref{fig:sn}. The region where the nulling process was performed is outlined by the dashed white line and spans from $5-22.5~\lambda/D$. An image without the nulling applied is shown in the top panel of Fig.~\ref{fig:sn} for comparison. This result was obtained on the $2^{nd}$ of June $2014$ in favorable seeing conditions (seeing $<0.7"$). The nulling process reduced the average flux over the entire controlled area by $30\%$ and by $58\%$ in the region between $5-12~\lambda/D$, where the nulling was most effective. With better wavefront correction and the use of a coronagraph, the improvement in the contrast will grow. 

The current limitations to achieving high-quality speckle nulling on-sky are: wavefront correction, readout noise and loop speed. As high sensitivity cameras in the NIR are currently limited in regards to maximum frame rate (the Axiom cameras used are amongst the fastest commercially available at the time of writing), it is not possible for the active speckle nulling algorithm to pursue atmospherically induced  speckles as they change from frame-to-frame. Hence, the current implementation of speckle nulling on SCExAO aims at removing the static and quasi-static speckles induced by diffraction from the secondary and spiders as well as optical aberrations. For this reason it is important to have a high level of atmospheric wavefront correction on-sky so that the persistent speckles due to static and quasi-static aberrations can be easily identified. As the speckles are $\sim1000\times$ fainter than the PSF core, a magnitude limit for speckle nulling of $3$--$4$ in the H-band is imposed by the current cameras used due to the high readout noise ($114e^{-}$). This places a severe limitation on potential targets of scientific interest. To alleviate these issues SCExAO is acquiring a Microwave Kinetic Inductance Detector (MKID) which is a photon counting, energy discriminating NIR array~\citep{mazin12}. The MKID array will offer almost no readout noise or dark current and is capable of high frame rates ($>1$~kHz). This enables speckle nulling to be performed on fainter more scientifically relevant targets and for non-common speckles due to chromatic dispersion in the atmosphere to be corrected for the first time, allowing for a significant improvement in detectability of faint companions. As the developmental time for the MKID array is several years, speckle control is being tested with a SAPHIRA (SELEX) array of avalanche photodiodes in the interim~\citep{at14}.  

In contrast to speckle suppression, the addition of artificial speckles to the focal plane image can be utilized for precision astrometry when the on-axis starlight has been suppressed post-coronagraph. Further, by modulating the phase of the speckles during an exposure they can be made incoherent with the speckles in the halo offering superior astrometric performance. In addition, by carefully calibrating the flux ratio between the PSF core and speckles it is also possible to use these speckles for photometry and hence retrieving the contrast of companions as well~\citep{martinache14b}. As opposed to diffractive grids utilized by other high-contrast imagers~\citep{wang14}, the adaptive nature of the DM allows the speckle position, and brightness to be carefully tailored to each science case.

\subsection{Coronagraphs}\label{sec:cor}
The advanced wavefront control techniques utilized on SCExAO build the foundation for high-contrast ($10^{-5}$--$10^{-6}$) imaging of faint companions with the onboard coronagraphs. The coronagraphs available in SCExAO are listed in Table~\ref{tab:corona}. The performance of the coronagraphs used in SCExAO is limited by the level of wavefront control achieved. The PIAA/PIAACMC and the Vortex offer the lowest IWA and highest throughput but are more sensitive to wavefront error. On the other hand the shaped pupil has a larger IWA and lower throughput but is less sensitive to residual wavefront error. Hence, the coronagraphs available are designed to span a large range of residual wavefront error and should be chosen accordingly. 

\begin{deluxetable}{cccc}[h!]
\tabletypesize{\footnotesize}
\centering
\tablecaption{Details of SCExAO coronagraphs. \label{tab:corona}}
\tablehead{
\colhead{Coronagraph type}		& 	\colhead{Inner working angle } 	&	\colhead{Waveband(s)}        	\\  
							& 	\colhead{($\lambda/D$)}		&					}		\\
\startdata      
PIAA			  				& 	$1.5$					&	$y$-$K$					\\
PIAACMC						& 	$0.8$					&	$y$-$K$					\\
Vortex						&	$2$						&	$H$						\\
MPIAA + Vortex				& 	$1$						&	$H$						\\
MPIAA + 8 Octant 				&   	$2$						&	$H$						\\     
4 quadrant 					&   	$2$						&	$H$						\\     
Shaped pupil 					&   	$3$						&	$y$-$K$					\\ 
\enddata
\end{deluxetable}

\begin{figure}[t!]
\centering 
\includegraphics[width=0.99\linewidth]{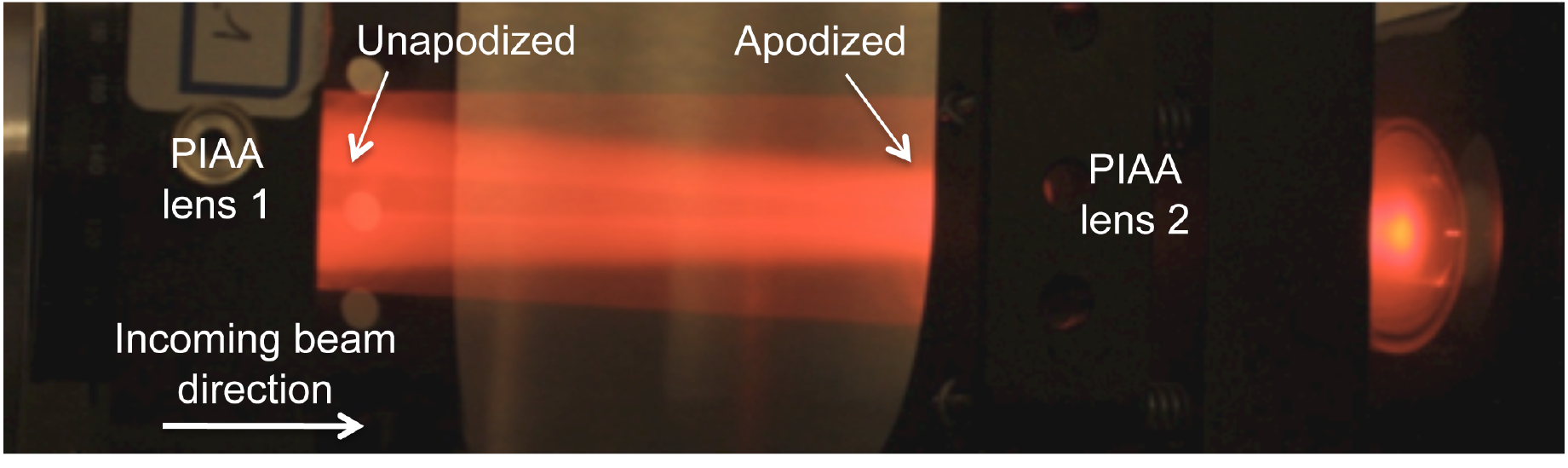}
\caption{\footnotesize Picture of a visible beam being apodized by the conventional PIAA lenses in the laboratory (side view). The image is taken between the two lenses. The beam enters from the left of the image where the intensity across the beam is uniform except for the faint part in the middle of the beam (behind the secondary) and at the right hand side of the image, the beam is concentrated in the middle (i.e. apodized).}
\label{fig:piaa}
\end{figure}

\begin{figure}[t!]
\centering 
\includegraphics[width=0.99\linewidth]{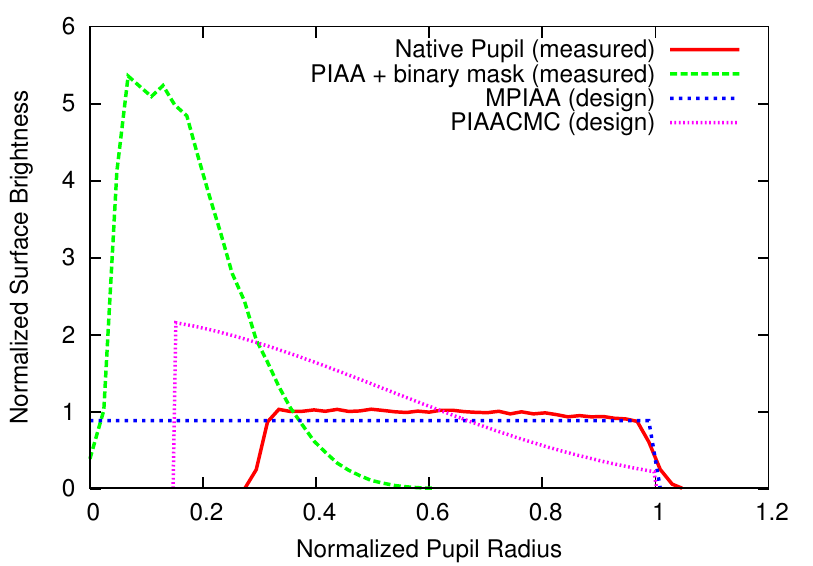}
\caption{\footnotesize The figure shows a comparison of the radial apodization profiles of the various PIAA lenses as compared to the Subaru telescope pupil. The PIAA (green line) pulls the light inwards most aggressively, almost entirely removing the secondary and softens the edges of the beam. The PIAACMC (pink line) pulls some of the light inwards but does not entirely remove the presence of the secondary or soften the edges completely. The MPIAA (blue line) removes the presence of the secondary but does not soften the edges at all. }
\label{fig:apo}
\end{figure}

The two key coronagraphs are the Phase induced amplitude apodization (PIAA) and the vector vortex types. PIAA refers to the act of remapping a flat-top pupil to a soft edged pupil in order to remove the diffraction features associated with a hard edged aperture (i.e. Airy rings)~\citep{guyon03}. These diffraction features make it difficult to suppress all of the light via a coronagraphic mask in the focal plane without blocking a close, faint companion. A combination of aspheric lenses are used to achieve the remapping in SCExAO and are referred to as PIAA lenses. SCExAO offers several types of remapping lenses. The first type is referred to as the conventional PIAA design and was presented in \cite{lozi09}. Conventional PIAA lenses offer the most aggressive remapping, eliminating the secondary and converting the post-PIAA pupil into a prolate spheroid (a near-Gaussian which is finite in extent). An image depicting the remapping process between the two PIAA lenses in the laboratory is shown in Fig.~\ref{fig:piaa}, while a radial profile of the apodization function is shown in Fig.~\ref{fig:apo}. 

To complete the softening of the edges of the beam a binary mask is used which has a radially variant attenuation profile. Note that the binary mask is used, to reduce the demand on the curvature changes across the aspheric surfaces of the PIAA lenses and it is possible to eliminate it at the expense of increased manufacturing complexity of the PIAA lenses. As outlined in \cite{guyon03}, once the on-axis star has been suppressed with a focal plane mask, it is important to reformat the pupil to its original state in order to preserve the field-of-view. This can be done by using another set of PIAA lenses in reverse and the ensemble of lenses are referred to as the inverse PIAA lenses (IPIAA). The position of the PIAA and inverse PIAA lenses can be seen in Figure~\ref{fig:scexao}. Due to the low material dispersion of CaF$_{2}$, the conventional PIAA lens design used in SCExAO is achromatic across the NIR (y-K bands). However, an appropriate focal plane mask must be chosen to achieve this. If an ordinary opaque mask is used, then the size of the mask is wavelength dependent, and so is the IWA. To circumvent this issue, SCExAO uses focal plane masks that consist of a central cone surrounded by a ring of pits periodically positioned around the cone, made from a transmissive material on a substrate which refracts rather than reflects the on-axis star light~\citep{newman15}. In this way a variable focal plane mask which is achromatic across H-band can be achieved. In addition, since the light is strongly diffracted outwards by the focal plane masks, it can be redirected towards the LLOWFS via a reflective Lyot mask~\citep{singh14}. 


Despite the fact that the conventional PIAA offered in SCExAO has an IWA of $1.5~\lambda/D$, due to the aggressive remapping which causes an abrupt phase step in the central part of the beam post-PIAA, the contrast at $1.5~\lambda/D$ is limited to $1\times10^{-5}$ and is very sensitive to tip/tilt. To alleviate these issues a modified version of the PIAA coronagraph can by used. It is referred to as the PIAA complex mask coronagraph (PIAACMC) and is outlined in greater detail in~\cite{guyon10a}. The major difference is that the PIAA lenses used for the PIAACMC are less aggressive which means the remapped pupil has soft edges but the secondary is still present as shown in Fig.~\ref{fig:apo}. The lenses themselves are in the same mounts as those for the PIAA so they can be replaced on the fly. The focal plane mask is now replaced with a partially transmissive, phase shifting mask which is manufactured via electron beam etching. The IWA of the coronagraph can be tuned by varying the opacity of the focal plane mask and in the limit when the mask is fully transmissive, the IWA is minimized at the expense of sensitivity to tip/tilt. Nonetheless, the PIAACMC offers a contrast of $~1\times10^{-6}$ at $1~\lambda/D$, is less sensitive to tip/tilt than the PIAA and is fully achromatic from y-K band. It is scheduled to be installed and commissioned in the near future.

A third and final type of PIAA is used to remap the pupil into a flat top without a central obstruction for an $8$-octant coronagraph which is discussed in the following section~\citep{osh14}. The lenses are referred to as MPIAA lenses and reside in the same mounts as the other two. Despite the remapping these optics are not apodizers as the pupil retains its hard edge post remapping. A comparison of the various apodization schemes is shown in Fig.~\ref{fig:apo}.

Other coronagraphs include the vortex~\citep{mawet09,mawet10}, $4$-quadrant~\citep{rouan00}, $8$-octant~\citep{mura10} and shaped pupil~\citep{carlotti12} versions. The vortex, $4$-quadrant and $8$-octant coronagraphs are phase-mask coronagraphs as opposed to occulting coronagraphs and consist of two primary elements; a focal plane and Lyot stop mask. All focal plane masks are situated in a wheel in the focal plane while the Lyot stop masks are located in the Lyot mask wheel and the positions of both are shown in Fig.~\ref{fig:scexao}. 

The vortex coronagraph in SCExAO uses a H-band optimized, topographic charge $2$ focal plane mask. This mask is constructed from a birefringent liquid crystal polymer material, i.e. a waveplate where the optical axis (fast axis) orientation is spatially dependent and in this case a function of the azimuthal coordinate~\citep{mawet09}. Although an IWA of $0.9~\lambda/D$ is achievable with an unobstructed pupil with non-manufacturing defects, it is limited to $2.0~\lambda/D$ with the pupil geometry at the Subaru Telescope. However, the vector vortex on SCExAO could be used in conjunction with the MPIAA lenses to circumvent this issue and regain the inherent IWA. The vector vortex mask is more achromatic than a scalar mask and hence can operate across the full H-band~\citep{mawet09}. As the nulling process is based on interference of light from different parts of the mask, best performance is achieved with higher Strehl ratios and stable centering of the PSF on the mask ($5$~mas tip/tilt error or below in this case). 

The $8$-octant coronagraph focal plane mask employed on SCExAO is based on photonic crystal technology~\citep{mura10}. It consists of $8$ triangular segments that comprise half-wave plates where the optical axes of a given segment is always orthogonal to its two nearest neighbors. This creates a $\pi$ phase shift between adjacent segments for the transmitted beam which destructively interferes in the reimaged focal plane to null out the on-axis star. The $8$ octant itself is not achromatic but broadband operation can be realized by placing a polarizer and analyzer before and after the mask respectively~\citep{mura08} (note this is also true for the vortex coronagraph). This coronagraph exploits the pupil-reformatting MPIAA lenses described above to achieve an IWA of $\sim2~\lambda/D$ and offers very high-contrasts at these angular scales. Similar to the vector vortex it is also sensitive to tip/tilt and hence active control is preferred mode of operation.

The $4$-quadrant focal plane mask is a scalar mask which consists of segments that phase shift the light by $\pi$ with respect to the neighboring segments. Although a perfectly manufactured $4$-quadrant mask could offer an IWA of as low as $1~\lambda/D$ if used in conjunction with the MPIAA lenses mentioned above, the mask in SCExAO has manufacturing defects and so can not achieve such performance. The $4$-quadrant in SCExAO is a prototype which serves its purpose for internal testing only.

The Lyot stop masks for the vector vortex, $4$-quadrant and $8$~octant coronagraphs are designed to reflect rejected light towards the LOWFS camera for fine tip/tilt guiding which is discussed in the subsequent section. The vector vortex and $4$-quadrant Lyot stop masks consist of a replica mask to the Subaru Telescope pupil  geometry with slight modifications. Both masks have a slightly oversized secondary and spiders for better rejection, however, the $4$-quadrant has a square secondary instead of a circular one. On the other hand, as the secondary is eliminated thanks to the MPIAA lenses, then the Lyot stop for the $8$-octant is simply a slightly undersized circular aperture. 
 
Finally, shaped pupil coronagraphs can also be tested on SCExAO. These coronagraphs are located in the pupil plane mask wheel and any focal or Lyot plane masks required are placed in the appropriate wheel. For further details please see~\cite{carlotti12}.

\subsection{Lucky Fourier Imaging}\label{sec:LFI}
\begin{figure}[t!]
\centering 
\includegraphics[width=0.70\linewidth]{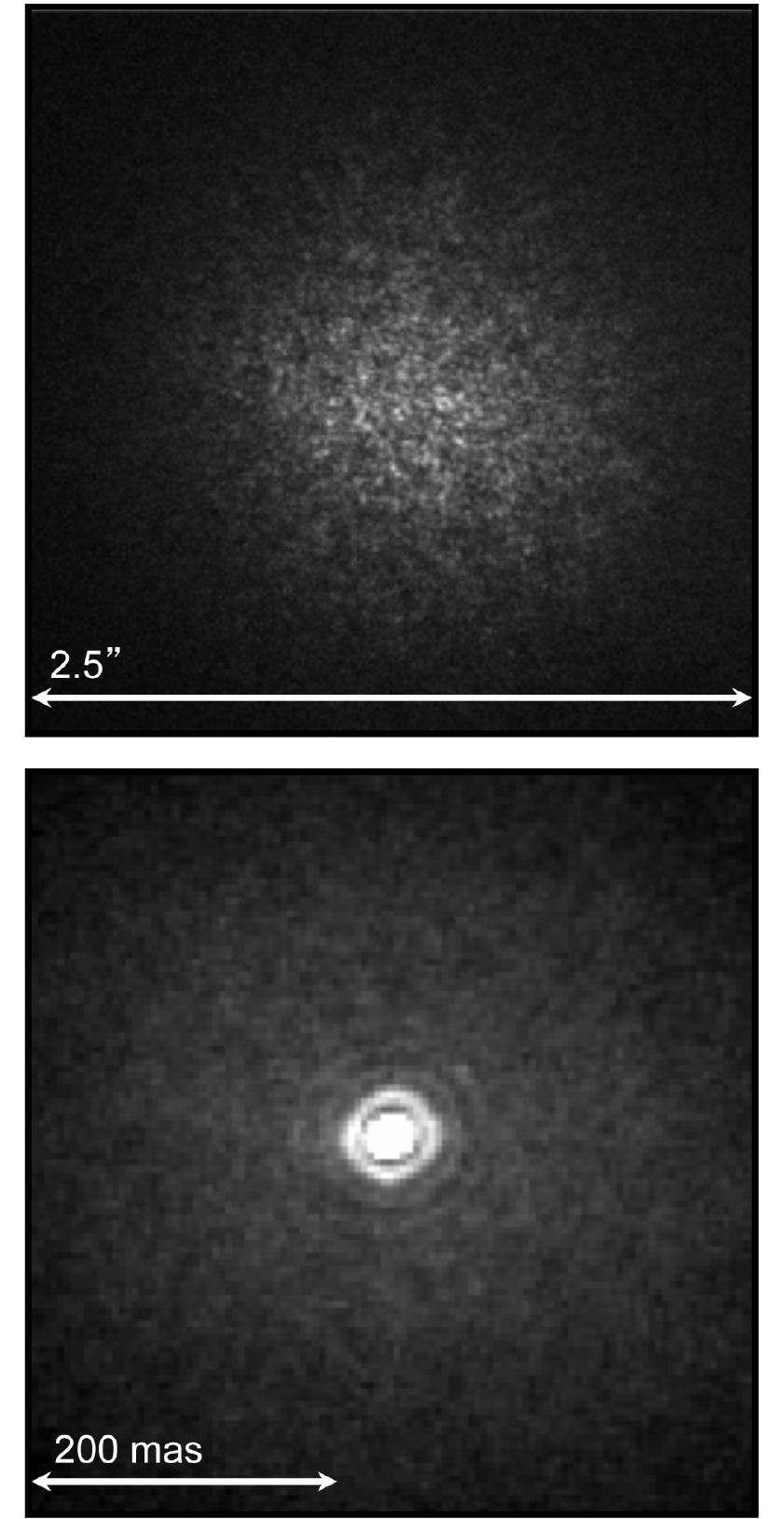}
\caption{\footnotesize Top image: Vega in 2" seeing at $680$~nm. Bottom image: Synthesized image of Vega at $680$~nm in the presence 2" seeing. Image was reconstructed from a $1\%$ selection of Fourier components across the $10^{4}$ frames collected. A diffraction-limited PSF with a FWHM of $17$~mas is obtained post-reconstruction (note: the scale of the bottom image differs from the top one). The data was acquired on the $5^{th}$ and $6^{th}$ of February, $2012$.}
\label{fig:lucky}
\end{figure}

An important element of all adaptive optics systems is a real-time PSF monitoring camera. This is depicted as the Lucky imaging camera in Fig.~\ref{fig:scexao}, the specifications of which are listed in Table~\ref{tab:detectors}. Currently a narrowband of light ($\sim30-50$~nm) is steered towards this camera from the pupil plane masks of VAMPIRES and the PSF imaged. The camera runs at a high frame rate, sub-framed and collects images rapidly which are primarily used for monitoring the PSF. The frames can subsequently also be used for traditional lucky imaging. However, a more advanced version of this technique named Lucky Fourier Imaging is commonly utilized~\citep{garrel12a}. The technique relies on looking for the strongest Fourier components of each image, and then synthesizing a single image with the extracted Fourier information. In this way diffraction-limited images at $680$~nm of targets like Vega (bottom image in Fig.~\ref{fig:lucky}) and Betelgeuse have been synthesized in $2"$ seeing~\citep{garrel12b}. 
This is clearly an extremely powerful tool and we propose to advance this imaging capability by adding multiple spectral channels.

\subsection{Fiber injection unit}\label{sec:SMI}
In addition to direct imaging, long baseline interferometry and high precision radial velocity both stand to gain significantly from a stable and $90\%$ Strehl PSF on an $8$-m class telescope. For example, long baseline interferometers like the Optical Hawaiian Array for Nano-radian Astronomy (OHANA) combine beams from multiple telescopes once it has been transported to the combination room via single-mode optical fibers~\citep{woillez04}. However, coupling efficiently into single-mode fibers is no mean feat, but with access to a stable PSF with $90\%$ Strehl, it can be achieved. Indeed this is already being exploited for the purposes of nulling interferometry on P$1640$~\citep{serebyn2010b}. Once the light has been coupled into a single-mode fiber, it could be used as an alternative feed for a conventional multimode fiber-fed spectrograph. The non-temporally and spatially varying PSF provided by a single-mode fiber can be used in precision radial velocity measurements to eliminate modal-noise, a limiting factor in achieving high precisions. For these reasons we are developing a single-mode injection unit on the SCExAO platform. 

To inject light into the fiber, it is tapped off with a retractable dichroic on its way to focus after OAP$2$ (see Fig.~\ref{fig:scexao}). A dichroic which reflects y, J and H-short bands is currently used for this. An achromatic lens ($f=10$~mm) is used to adjust the $F/\#$ of the beam before it is injected into the fiber which sits atop a stage. The $5$-axis stage allows for XYZ translation via precise stepper motor actuators and course alignment of tip/tilt. The stage can be scanned through focus to maximize coupling into the fiber. A further advantage of implementing such a module on SCExAO is that we can use the conventional PIAA lenses to more closely match the intensity distribution of the collection fiber and hence boost the coupling to a theoretical value of $100\%$. To make this useful on-sky, this injection system relies on the PyWFS delivering a high Strehl beam. The fine tip/tilt control is provided by the LLOWFS by using the transmitted H-band light. This unit is currently operational and may be utilized by instruments such as the high precision spectrograph IRD for the survey it will undertake~\citep{tamura12}. In addition, by developing such a unit, it becomes possible to exploit numerous other photonics technologies on-sky~\citep{cvetojevic12,marien12}. The injection unit is described in more detail in~\cite{Jovanovic14b}.

\section{Commissioning status and future extensions}\label{sec:future}
SCExAO is clearly a complex instrument with modules at various stages of commissioning. Table~\ref{tab:commission} summarizes the commissioning status/plan of the various modes of operation and modules of SCExAO. 
\begin{deluxetable}{llc}[ht!]
\tabletypesize{\footnotesize}
\centering
\tablecaption{Commissioning status of SCExAO modes of \\operation and modules. \label{tab:commission}}
\tablehead{
\colhead{Module/mode}			& 	\colhead{Commissioning status}		   		\\ 
\colhead{}						&	\colhead{(completion date)}				}	\\
\startdata
\textbf{Wavefront control}			&		 									\\
LLOWFS						&	Complete									\\
SN							&	Complete									\\
PyWFS						&  	Partially complete (fall $2015$)					\\
PyWFS+LLOWFS				&  	Partially complete (fall $2015$)					\\
SN+PyWFS+LLOWFS			&     	Incomplete (late $2015$)			  			\\
							&											\\
\textbf{Coronagraphs}			&											\\
PIAA							&	Complete 									\\
Vortex						&	Complete									\\
4 quadrant					&	Complete									\\
Shaped pupil					&	Complete									\\	
PIAACMC						&	Incomplete (fall $2015$)						\\
MPIAA+Vortex					&	Incomplete (late $2015$)						\\
MPIAA+$8$ octant				&	Incomplete (fall $2015$)						\\	
							&											\\
\textbf{Visible imagers}			&											\\
VAMPIRES					&	Complete									\\  
Lucky imaging 					&      Partially complete (late $2015$)				\\
FIRST						&      Incomplete (late $2015$)						\\
							&											\\
NIR fiber injection 				&      Complete									\\ 
							&											\\
\textbf{NIR science imagers}		&											\\
CHARIS						&	Expected delivery early $2016$ (mid $2016$)		\\
MKID						&     	Expected delivery late $2016$ ($2017$)			\\ \enddata
\tablecomments{+ signifies that these modes/modules are operating in conjunction, SN-speckle nulling, PIAA will be replaced by the PIAACMC.}
\end{deluxetable}

Table~\ref{tab:commission} shows that the integral field spectrograph known as CHARIS will replace the HiCIAO imager from mid $2016$. This instrument segments the focal plane with an array of micro-lenses, before dispersing each PSF and then reimaging onto a detector (see Table~\ref{tab:detectors})~\citep{peters12}. This allows for spatially resolved spectral information albeit at low resolving powers. Such an instrument has three key advantages. Firstly, background stars in a given image can quickly be identified. Secondly, owing to the fact that the instrument operates over a very broadband (J-K bands), the presence of a planet can be inferred by detecting the fixed speckle within a spectral data cube. Finally, low resolution spectra of gas giants can be taken enabling the atmospheres of these planets to be constrained and better understood~\citep{michael12,brandt14}. Indeed, presently operating integral field spectrographs such as OSIRIS at the Keck Telescope, the units in P$1640$ and GPI have been used to characterize the atmospheres of known planetary systems like HR$8799$~\citep{bar11,Opp13, Ing14}.

Although not included in the table, it has been proposed to outfit the IR arm of SCExAO with a polarimetric mode of operation to study scattered dust in circumstellar disks. This mode known as polarization differential imaging (PDI) has been hugely successful for the HiCIAO imager~\citep{grady13}, and we aim to preserve this capability while offering a superior IWA. This IR polarimetric mode will complement the one of VAMPIRES offered in the visible, albeit on different spatial scales. 

\section{Summary}\label{sec:summary}
The SCExAO instrument is a versatile high-contrast imaging platform which hosts advanced wavefront control systems, IR coronagraphs and visible interferometers, that are ideal for imaging at $<3~\lambda/D$ (solar-system scales). The extreme adaptive optics system delivers the necessary wavefront correction to be able to push detection limits for ground-based observations at small angular separations and interferometer precisions. Such instruments will be critical to understanding the inner structure of circumstellar disks and planetary systems and how they form. In addition, they will provide the appropriate avenue to collecting spectra from planetary candidates and determining their physical properties for the first time. Further, the SCExAO platform is an ideal testbed for demonstrating and prototyping technologies for future ELTs and space-missions. SCExAO is the only high-contrast imager of its kind and will be uniquely positioned to contribute to exoplanetary science.

\section{Appendix}~\label{sec:appendix}
\subsection{Off-axis parabolic mirrors}
All OAPs in the SCExAO instrument were manufactured via diamond turning of aluminum and overcoating with gold, are $50$~mm in diameter and were designed to work at a nominal off-axis angle of $17^{\circ}$. OAPs $1,3,4, \&$ $5$ have a $f=255$~mm while OAP$2$ has a $f=519$~mm. Although, there is no data on the wavefront error of these optics, each optic was initially used to form an image in the visible and it was determined from this that the RMS wavefront error was $<\lambda/20$ at $630$~nm over a $20$~mm beam size corresponding to that which is used in SCExAO.

\subsection{Internal NIR camera lenses}
The focusing lenses for the science camera include a $f=150$ and $50$~mm converging, achromatic doublet which are AR-coated for the NIR region ($1-1.65~\mu$m). The distance between the lenses is set to be just larger than the sum of their focal lengths so that a slow beam is formed ($F/\#=65$).

\subsection{Deformable mirror environmental controls}
An interlock system which monitors the environmental conditions in the DM chamber (pressure, humidity) was put in place to prolong the life of the DM as they are known age rapidly in high humidity environments~\citep{shea04}. A low pressure regulator (Fairchild-M$4100$) was used to offer fine control of the injected dry air pressure to the DM chamber at the $<1$~psi level (with respect to ambient). The pressure is set to $0.4$~psi above ambient when operating and monitored by a precise pressure sensor (FESTO - SDE$1$). Such low-pressure differentials are used so that the chamber does not deform significantly and hence induce any extra errors to the wavefront. A $1$~psi pressure differential relief valve is used as a hard limit in case of over pressure in the circuit. The humidity in the circuit is measured with a moisture probe (Edgetech - HT$120$). The alarms for both the pressure and humidity sensors are used to control the power supplied to the DM electronics. When the humidity is below $15\%$ and the pressure between $0.2-0.8$~psi, the DM electronics will be powered and the actuators can be driven. However, if the humidity rises above $15\%$ and/or the pressure goes above $0.8$~psi or below $0.2$~psi, then the alarms on the sensors will trigger a relay switch, to which the DM electronics are connected to, to trip and cut the power to the DM. As a final level of reassurance, a low flow rate ($250$~mL/min) flow controller is connected to the end of the line to insure that there is a very slow flow over the DM membrane and no turbulence in the chamber. A rapid flow could tear the thin silicon membrane ($3~\mu$m) and or cause turbulence in the chamber which would be equivalent to seeing. The window to the chamber is $50$~mm in diameter, made from a $12$~mm thick piece of IR fused silica which is AR-coated across the operating range.

\acknowledgments
We are grateful to B. Elms for his contributions to the fabrication of parts for the SCExAO rebuild. The SCExAO team thanks the Subaru directorate for funding various grants to realize and develop the instrument.

\textit{Facilities: Subaru.}


\begin{thebibliography}{}
\bibitem[Artigau (2008)]{artigau08} Artigau, E., Biller, B. A., Wahhaj, Z., Hartung, M., Hayward, T. L., Close, L. M., Chun, M. R., Li, M. C., Trancho, G., Rigaut, F., Toomey, D. W., \& Ftaclas, C., 2008, Proc. of the SPIE, 70141Z
\bibitem[Atkinson \it{et al.} (2014)]{at14} Atkinson, D., Hall, D., Baranec, C., Baker, I., Jacobson, S., \& Riddle, R., $2014$, Proc. of the SPIE, 915419A
\bibitem[Barman \it{et al.} (2011)]{bar11} Barman, T. S., Macintosh, B., Konopacky, Q. M., \& Marois, C., $2011$, \apj, 733, 65B. 
\bibitem[Beuzit \it{et al.} (2008)]{bez08} Beuzit, J.-L., Feldt, M., Dohlen, K., Mouillet, D., Pugeta, P., Wildi, F., Abee, L., Antichif, J., Baruffolof, A., Baudozg, P., Boccaletti, A., Carbillet, M., Charton, J., Claudi, R., Downing, M., Fabron, C., Feautrier, P., Fedrigo, E., Fusco, T., Gach, J.-L., Gratton, R., Henning, T., Hubin, N., Joos, F., Kasper, M., Langlois, M., Lenzen, R., Moutou, R., Pavlov, A., Petit, C., Pragt, J., Raboua, P., Rigal, F., Roelfsema, R., Rousset, G., Saisse, M., Schmid, H.-M., Stadler, E., Thalmann, C., Turatto, M., Udry, S., Vakili, F., and Waters, R., 2008, Proc. of SPIE, 701418
\bibitem[Borde \& Traub (2006)]{borde06} Borde, P. J., \& Traub, W. A., 2006, \apj, 638, 488B
\bibitem[Borucki \it{et al.} (2010)]{borucki10} Borucki, B., Koch, D., Basri, G., Batalha, N., Brown, T., Caldwell, D., Caldwell, J., Christensen-Dalsgaard, J., Cochran, W. D., DeVore, E, Dunham, E. W., Dupree, A. K., Gautier, T. N., III, Geary, J. C., Gilliland, R., Gould, A., Howell, S. B., Jenkins, J. M., Kondo,  Latham, D. W., Marcy, G. W., Meibom, S., Kjeldsen, H., Lissauer, J. J., Monet, D. G. Morrison, D., Sasselov, D., Tarter, J., Boss, A., Brownlee, D., Owen, T., Buzasi, D., Charbonneau, D., Doyle, L., Fortney, J., Ford, E. B., Holman, M. J., Seager, S., Steffen, J. H., Welsh, W. F., Rowe, J., Anderson, H., Buchhave, L., Ciardi, D., Walkowicz, L., Sherry, W., Horch, E., Isaacson, H., Everett, M. E., Fischer, D., Torres, G., Johnson, J. A., Endl, M., MacQueen, P., Bryson, S. T., Dotson, J., Haas, M., Kolodziejczak, J., Cleve, J. V., Chandrasekaran, H., Twicken, J. D., Quintana, E. V., Clarke, B. D., Allen, C., Li, J., Wu, H., Tenenbaum, P., Verner, E., Bruhweiler, F., Barnes, J., and Prsa A, 2010, Science, 327, 977
\bibitem[Brandt \it{et al.} (2014)]{brandt14} Brandt, T., McElwain, M., Janson, M., Knapp, G., Mede, K., Limbach, M., Groff, T., Burrows, A., Gunn, J., Guyon, O., Hashimoto, J., Hayashi, M., Jovanovic, N., Kasdin, J. Kuzuhara, M., Lupton, R., Martinache, F., Sorahana, S., Spiegel, D., Takato, N., Tamura, M., Turner, E., Vanderbei, R., \& Wisniewski, J., $2014$, Proc. of SPIE, 9148-49
\bibitem[Buton \it{et al.} (2012)]{buton12} Buton, C., \textit{et al.} $2012$, A\&A, 549A, 5B
\bibitem[Carlotti \it{et al.} (2012)]{carlotti12} Carlotti, A., Kasdin, N. J., Martinache, F., Vanderbei, R., J., Young, E. J., Che, G., Groff, T. D., \& Guyon, O., 2012, Proc. of the SPIE, 84463C.
\bibitem[Charbonneau (2001)]{char01} Charbonneau, D., Brown, T. M., Noyes, R. W., and Gillilna, R. L., 2001, \apj, 568, 377
\bibitem[Clergeon (2014)]{clergeon14} Clergeon, C., PhD dissertation, 2014
\bibitem[Close \it{et al.} (2013)]{close13} Close, L. M., Males, J. R., Morzinski, K., Kopon, D., Follette, K., Rodigas, T. J., Hinz, P., Wu, Y.-L., Puglisi, A., Esposito, S., Riccardi, A., Pinna, E., Xompero, M., Briguglio, R., Uomoto, A., \& Hare, T., $2013$, \apj,774, 94C
\bibitem[Codana \& Kenworthy (2013)]{codona13} Codana, J. L., \& Kenworthy, M., 2013, \apj, 767, 100C
\bibitem[Cvetojevic \it{et al.} (2012)]{cvetojevic12} Cvetojevic, N., Jovanovic, N., Betters, C., Lawrence, J. S., Ellis, S., Robertson, G., Bland-Hawthorn, J., $2012$, A\&A, 544, L1
\bibitem[Dekany \it{et al.} (2013)]{dekany13} Dekany, R., Roberts, J., Burruss, R., Antonin, T., Baranec, C., Guiwits, S., Hale, D., Angione, J., Trinh, T., Zolkower, J., Shelton, C., Palmer, D., Hemming, J., Croner, E., Troy, M., McKenna, D., Tesch, J., Hildebrandt, S., Milburn, J., $2013$, \apj, 776, 130D
\bibitem[Esposito \it{et al.} (2010)]{esp10} Esposito, S., Riccardi, A., Fini, L., Puglisi, A. T., Pinna, E., Xompero, M., Briguglio, R., Quirós-Pacheco, F., Stefanini, P., Guerra, J. C., Busoni, L., Tozzi, A., Pieralli, F., Agapito, G., Brusa-Zappellini, G., Demers, R., Brynnel, J., Arcidiacono, C., \& Salinari, P., $2010$, Proc. of SPIE, 7736-09F.
\bibitem[Esposito \it{et al.} (2011)]{esp11} Esposito, S., Riccardi, A., Pinna, E., Puglisi, A., Quirós-Pacheco, F., Arcidiacono, C., Xompero, M., Briguglio, R., Agapito, G., Busoni, L., Fini, L., Argomedo, J., Gherardi, A., Brusa, G., Miller, D., Guerra, J. C., Stefanini, P., \& Salinari, P., $2011$, Proc. of SPIE, 8149-02E
\bibitem[Garrel \it{et al.} (2012)]{garrel12a} Garrel, V., Guyon, O., \& Baudoz, P., \pasp, 124, 861
\bibitem[Garrel (2012)]{garrel12b} Garrel, V., PhD dissertation, 2012
\bibitem[Give'on \it{et al.} (2007)]{giveon07} Give'on, A., Kern, B., Shaklan, S., Moody, D. C., \& Pueyo, L., 2007, Proc. of SPIE, 66910A
\bibitem[Gozdziewski \& Migaszewski (2014)]{goz14} Gozdziewski, K., \& Migaszewski, C., $2014$, \mnras, 440, 3140.
\bibitem[Grady \it{et al.} (2013)]{grady13} Grady, C. A. Muto, T., Hashimoto, J., Fukagawa, M.,  Currie, T., Biller, B., Thalmann, C., Sitko, M. L., Russell, R., Wisniewski, J., Dong, R., Kwon, J., Sai, S., Hornbeck, J., Schneider, G., Hines, D., Martín, A. M., Feldt, M., Henning, Th., Pott, J.-U., Bonnefoy, M., Bouwman, J., Lacour, S., Mueller, A., Juhász, A., Crida, A., Chauvin, G., Andrews, S., Wilner, D., Kraus, A.,  Dahm, S., Robitaille, T., Jang-Condell, H., Abe, L., Akiyama, E., Brandner, W., Brandt, T., Carson, J., Egner, S., Follette, K. B., Goto, M., Guyon, O., Hayano, Y., Hayashi, M., Hayashi, S., Hodapp, K., Ishii, M., Iye, M., Janson, M., Kandori, R., Knapp, G., Kudo, T., Kusakabe, N., Kuzuhara, M., Mayama, S., McElwain, M., Matsuo, T., Miyama, S., Morino, J.-I., Nishimura, T., Pyo, T.-S., Serabyn, G., Suto, H., Suzuki, R., Takami, M., Takato, N., Terada, H., Tomono, D., Turner, E., Watanabe, M., Yamada, T., Takami, H., Usuda, T., and Tamura, M., 2013, \apj,762, 48
\bibitem[Guyon (2003)]{guyon03} Guyon, O., 2003, A\&A, 404, 379
\bibitem[Guyon (2005)]{guyon05} Guyon, O., 2005, \apj, 629, 592
\bibitem[Guyon \it{et al.} (2009)]{guyon09} Guyon, O., Matsuo, T., \& Angel, R. 2009, \apj, 693, 75
\bibitem[Guyon \it{et al.} (2010)]{guyon10a} Guyon, O., Martinache, F., Belikov, R., and Soummer, R.  2010, \apjs, 190, 220
\bibitem[Guyon \it{et al.} (2010)]{guyon10b} Guyon, O., Pluzhnik, E., Martinache, F., Totems, J., Tanaka, S., Matsuo, T., Blain, C., \& Belikov, R., 2010, \pasp, 122, 71G
\bibitem[Han \it{et al.} (2014)]{han14} Han, E., Wang, S., Wright, J. T., Feng, Y., K., Zhao, M., Fakhouri, O., Brown, J., I., \& Hancock, C., $2014$, \pasp, 126, 827
\bibitem[Hardy (1998)]{hardy98} Hardy, J. W., $1998$, Adaptive optics for astronomical telescopes, Oxford University Press, New York
\bibitem[Helmbrecht (2011)]{helm11} Helmbrecht, M. A., Min H., Carl K., \& Marc B., $2011$, Proc. of SPIE 793108.
\bibitem[Huby \it{et al.} (2012)]{huby12} Huby, E., Perrin, G., Marchis, F., Lacour, S., Kotani, T., Duchene, G., Choquet, E., Gates, E. L., Woillez, J. M., Lai, O., Fedou, P., Collin, C., Chapron, F., Arslanyan, V., and Burns, K. J., 2012, A\&A, 541A, A55
\bibitem[Huby \it{et al.} (2013)]{huby13} Huby, E., Duchene, G., Marchis, F., Lacour, S., Perrin, G., Kotani, T.,  Choquet, E., Gates, E. L., Lai, O., Allard, F., 2013, A\&A, 560, A113
\bibitem[Ingraham \it{et al.} (2014)]{Ing14} Ingraham, P., \textit{et al.}, $2014$, \apj, 794L, 15I.  
\bibitem[Jovanovic \it{et al}. (2014a)]{Jovanovic14a} Jovanovic, N., Guyon, O., Martinache, F., Clergeon, C., Singh, G., Kudo, T., Newman, K., Kuhn, J., Serabyn, E., Norris, B., Tuthill, P., Stewart, P., Huby, E., Perrin, G., Lacour, S., Vievard, S., Murakami, N., Fumika, O., Minowa, Y., Hayano, Y., White, J., Lai, O., Marchis, F., Duchene, G., Kotani, T., Woillez, J., $2014$, Proc. of SPIE, 9147-1Q.
\bibitem[Jovanovic \it{et al}. (2014b)]{Jovanovic14b} Jovanovic, N., Guyon, O, Martinache, F., Schwab, C., \& Cvetojevic, N., $2014$, Proc. of SPIE, 9147, 9147-287 
\bibitem[Kraus \& Ireland (2012)]{kraus12} Kraus A., \& Ireland M. J., $2012$, \apj, $745$, $1$
\bibitem[Lafreniere \it{et al}. (2009)]{Laf09} Lafreniere, D., Marois, C., Doyon, R., \& Barman, T., $2009$, \apj, 694L, 148L. 
\bibitem[Lagrange \it{et al}. (2009)]{Lag09} Lagrange, A.-M., Gratadour, D., Chauvin, G., Fusco, T., Ehrenreich, D., Mouillet, D., Rousset, G., Rouan, D., Allard, F., Gendron, É., Charton, J., Mugnier, L., Rabou, P., Montri, J., \& Lacombe, F., $2009$, A\&A, 493, 21L. 
\bibitem[Lenzen \it{et al}. (2003)]{Lenzen03} Lenzen, R., Hartung, M., Brandner, W., Finger, G., Hubin, N., Lacomber, F., Lagrange, A.-M., Lehnert, M., Moorwood, A., \& Mouillet, D., $2003$, Proc. of SPIE, 4841-944L
\bibitem[Leon-Saval \it{et al.} ($2013$)]{saval2013} Leon-Saval, S. G., Argyros, A., \& Bland-Hawthorn, J., $2013$, Nanophot., 2, 429. 
\bibitem[Lord (1992)]{lord92} Lord, S. D., $1992$, NASA Technical Memorandum $103957$
\bibitem[Lozi \it{et al.} (2009)]{lozi09} Lozi, J., Marinache, F., and Guyon, O.,  2009, \pasp, 121, 1232
\bibitem[Macintosh (2014)]{macintosh14} Macintosh, B. et. al., $2014$, PNAS, 111, 12661.
\bibitem[Malbet \it{et al.}(1995)]{malbet95} Malbet,  F., yu, J. W., \& Shao, M., 1995, \pasp, 107, 386M
\bibitem[Marien \it{et al.} (2012)]{marien12} Marien, G., N., Jovanovic, Cvetojevic, N., Williams, R., Haynes, R., Lawrence, Parker, Q., \& Withford, M., 2012, MNRAS, 421, 3641
\bibitem[Marois \it{et al.} (2008)]{marois08} Marois, C., Macintosh, B., Barman, T., Zuckerman, B.; Song, I., Patience, J., Lafrenière, D., \& Doyon, R., $2008$, Science, 322, 1348M. 
\bibitem[Martinache \it{et al.} (2012)]{martinache12} Martinache,  F., Guyon, O., Clergeon, C., and Blain, C., 2012, \pasp, 124, 1288
\bibitem[Martinache (2013)]{martinache13} Martinache,  F., 2013, \pasp, 125, 422M
\bibitem[Martinache \it{et al.} (2014a)]{martinache14a} Martinache,  F., Guyon, O., Jovanovic, N., Clergeon, C., Singh, G., Kudo, T., Currie, T., Thalmann, C., McElwain, M., \& Tamura, M., 2014, Proc. of the SPIE, 9148, 914870
\bibitem[Martinache \it{et al.} (2014b)]{martinache14b} Martinache,  F., Guyon, O., Jovanovic, N., Clergeon, C., Singh, G., Kudo, T., Currie, T., Thalmann, C., McElwain, M., \& Tamura, M., 2014, \pasp, 126, 565
\bibitem[Mawet \it{et al.} (2009)]{mawet09} Mawet, D., Serabyn, E., Liewer, K., Hanot, Ch., McEldowney, S., Shemo, D., \& O'Brien, N., 2009, Opt. Exp., 17, 1902
\bibitem[Mawet \it{et al.} (2010)]{mawet10} Mawet, D., Serabyn, E., Liewer, K., Burruss, R., Hickey, J., \& Shemo, D., 2010, \apj, 709, 53
\bibitem[Mayor \& Queloz (1995)]{mayor95} Mayor, M., and Queloz, D., 1995, Nature, 378, 355
\bibitem[Mazin \it{et al.}(2012)]{mazin12} Mazin, B. A., Bumble, B., Meeker, S. R., O?Brien, K., McHugh, S., \& Langman, E., $2012$, Opt. Exp., 20, 1503.
\bibitem[McElwain \it{et al.} (2012)]{michael12} McElwain, M. W., Brandt, T. D., Janson, M., Knapp, G. R., Peters, M. A., Burrows, A. S., Carlotti, A., Carr, M. A., Groff, T., Gunn, J. E., Guyon, O., Hayashi, M., Kasdin, N. J., Kuzuhara, M., Lupton, R. H., Martinache, F., Spiegel, D., Takato, N., Tamura, M., Turner, E. L., \& Vanderbei, R. J., 2012, Proc. of SPIE, 84469C
\bibitem[Minowa \it{et al.}(2010)]{minowa10} Minowa, Y., Hayano, Y., Oya, S., Watanabe, M., Hattori, M., Guyon, O., Egner, S., Saito, Y., Ito, M., Takamia, H., Garrel, V., Colley, S., Golota, T., \& Iye, M., 2010, Proc. of SPIE, 77363N
\bibitem[Murakami \it{et al.}(2008)]{mura08} Murakami, N., Uemura, R., Baba, N., Nishikawa, J., Tamura, M., Hashimoto, N., \& Abe, L., 2008, \pasp, 120, 1112
\bibitem[Murakami \it{et al.}(2010)]{mura10} Murakami, N., Nishikawa, J., Yokochi, K., Tamura, M., Baba, N., \& Abe, L., 2010, \apj, 714, 772
\bibitem[Muto \it{et al.}(2012)]{muto12} Muto, T., Grady, C. A., Hashimoto, J., Fukagawa, M., Hornbeck, J. B., Sitko, M., Russell, R., Werren, C., Cura, M., Currie, T., Ohashi, N., Okamoto, Y., Momose, M., Honda, M., Inutsuka, S., Takeuchi, T., Dong, R., Abe, L., Brandner, W., Brandt, T., Carson, J., Egner, S., Feldt, M., Fukue, T., Goto, M., Guyon, O., Hayano, Y., Hayashi, M., Hayashi, S., Henning, T., Hodapp, K. W., Ishii, M., Iye, M., Janson, M., Kandori, R., Knapp, G. R., Kudo, T., Kusakabe, N., Kuzuhara, M., Matsuo, T., Mayama, S., McElwain, M. W., Miyama, S., Morino, J.-I., Moro-Martin, A., Nishimura, T., Pyo, T.-S., Serabyn, E., Suto, H., Suzuki, R., Takami, M., Takato, N., Terada, H., Thalmann, C., Tomono, D., Turner, E. L., Watanabe, M., Wisniewski, J. P., Yamada, T., Takami, H., Usuda, T., Tamura, M., 2012, \apj~Letters, 748, L22
\bibitem[Newman \it{et al.}(2015)]{newman15} Newman, K., Guyon, O., Balasubramanian, K., Belikov, R., Jovanovic, N., Martinache, F., \& Wilson, D., $2015$, \pasp, 127, 437
\bibitem[Norris \it{et al.}(2012)]{norris12a} Norris, B., Tuthill, P. G., Ireland, M. J., Lacour, S., Zijlstra, A. A., Lykou, F., Evans, T. M., Stewart, P., Bedding, T. R., Guyon, O., \& Martinache, F.,   2012, Proc. of SPIE, 844503N.
\bibitem[Norris \it{et al.}(2012)]{norris12b} Norris, B., Tuthill, P. G., Ireland, M. J., Lacour, S., Zijlstra, A. A., Lykou, F., Evans, T. M., Stewart, \& P., Bedding,   2012, Nature, 484, 220
\bibitem[Norris \it{et al.}(2015)]{norris15} Norris, B., Schworer, G., Tuthill, P., Jovanovic, N., Guyon, O., Stewart, P., \& Martinache, F., $2015$, MNRAS, 447, 2894N.
\bibitem[Oppenheimer \it{et al.}(2013)]{Opp13} Oppenheimer, B. R., Baranec, C., Beichman, C., Brenner, D., Burruss, R., Cady, E., Crepp, J. R., Dekany, R., Fergus, R., Hale, D., Hillenbrand, L., Hinkley, S., Hogg, David W., King, D., Ligon, E. R., Lockhart, T., Nilsson, R., Parry, I. R., Pueyo, L., Rice, E., Roberts, J. E., Roberts, L. C., Jr., Shao, M., Sivaramakrishnan, A., Soummer, R., Truong, T., Vasisht, G., Veicht, A., Vescelus, F., Wallace, J. K., Zhai, C., Zimmerman, N., $2013$, \apj, 768. 24O.
\bibitem[Oshiyama \it{et al.}(2014)]{osh14} Oshiyama, F., Murakami, N., Guyon, O., Martinache, F., Baba, N., Matsuo, T., Nishikawa, J., \& Tamura, M., $2014$, \pasp, 126, 270 
\bibitem[Pepe \it{et al.} (2014)]{pepe14} Pepe, F., Ehrenreich, D., \& Meyer, M. R., $2014$, Nature, 513, 358.  
\bibitem[Perrin \it{et al.} (2006)]{perrin06} Perrin, G., Lacour, S., Woillez, J., \& Thiebaut, E., 2006, \mnras, 373, 747P
\bibitem[Peters \it{et al.} (2012)]{peters12} Peters, M. A., Groff, T., Kasdin, N. J., McElwain, M. W., Galvin, M., Carr, M. A., Lupton, R., Gunn, J. E., Knapp, G., Gong, Q., Carlotti, A., Brandt, T., Janson, M., Guyon, O., \bibitem[Poyneer \it{et al.} (2014)]{poy14} Poyneer, L. A., De Rosa, R. J., Macintosh, B., Palmer, D. W., Perrin, M. D., Sadakuni, N., Savransky, D., Bauman, B., Cardwell, A., Chilcote, J. K., Dillon, D., Gavel, D., Goodsell, S. J., Hartung, M., Hibon, P., Rantakyrö, F. T., Thomas, S., Veran, J.-P., $2014$, Proc. of SPIE, $9148$E, $0$KP.
\bibitem[Rouan \it{et al.} (2000)]{rouan00} Rouan, D., Riaud, P., Boccaletti, A., Clenet, Y., \& and Labeyrie, A., $2000$, \pasp, 112, 1479 
\bibitem[Serabyn \it{et al.} ($2010a$)]{serebyn2010a} Serabyn, E., Mawet, D., \& Burruss, R., $2010$, Nature, 464, 1018.
\bibitem[Serabyn \it{et al.} ($2010b$)]{serebyn2010b} Serabyn, E., Mennesson, B., Martin, S., Liewer, K., Mawet, D., Hanot, C., Loya, F., Colavita, M. M., \& Ragland, S., $2010$, Proc. of SPIE 77341E.
\bibitem[Shea \it{et al.}(2004)]{shea04} Shea, H. R., Gasparyan, A., Chan, H. B., Arney, S., Frahm, R. E., Lopez, D., Jin, S., \& McConnell, P., 2004, Trans. on Dev. \& Mat. Rel., 4, 198
\bibitem[Singh \it{et al.}(2014)]{singh14} Singh, G., Martinache, F., Baudoz, P., Guyon, O., Matsuo, T., Jovanovic, N., \& Clergeon, C.,  2014, \pasp, 126, 586
\bibitem[Singh \it{et al.}(2015)]{singh15} Singh, G., Lozi, J., Guyon, O., Jovanovic, N., Baudoz, P., Martinache, F., Kudo, T., Serabyn, E., \& Kuhn J., G., $2015$, in preparation.
\bibitem[Tamura \it{et al.} (2009)]{tamura09} Tamura, M., $2009$, Proc. of AIPC, 1158, 11T
\bibitem[Tamura \it{et al.} (2012)]{tamura12} Tamura, M., Suto, H., Nishikawa, J., Kotani, T., Sato, B., Aoki, W., Usuda, T., Kurokawa, T., Kashiwagi, K., Nishiyama, S., Ikeda, Y., Hall, D., Hodapp, K., Hashimoto, J., Morino, J., Inoue, S., Mizuno, Y., Washizaki, Y., Tanaka, Y., Suzuki, S., Kwon, J., Suenaga, T., Oh, D., Narita, N., Kokubo, E., Hayano, Y., Izumiura, H., Kambe, E., Kudo, T., Kusakabe, N., Ikoma, M., Hori, Ya., Omiya, M., Genda, H., Fukui, A., Fujii, Y., Guyon, O., Harakawa, H., Hayashi, M., Hidai, M., Hirano, T., Kuzuhara, M., Machida, M., Matsuo, T., Nagata, T., Ohnuki, H., Ogihara, M., Oshino, S., Suzuki, R., Takami, H., Takato, N., Takahashi, Y., Tachinami, C., \& Terada, H., 2012, Proc. of SPIE, 84461T
\bibitem[Tuthill \it{et al.} (2000)]{tuthill00} Tuthill, P., Monnier, J. D., Danchi W. C., Wishnow E. H., Haniff C. A., $2000$, \pasp, $112$, $555$
\bibitem[Tuthill \it{et al.} (2010)]{tuthill10} Tuthill, P., Lacour, S., Amico, P., Ireland, M., Norris, B., Stewart, P., Evans, T., Kraus, A., Lidman, C., Pompei, E., \& Kornweibel, N., $2010$, Proc. of the SPIE, 77351OT
\bibitem[Vogt \it{et al.} (2011)]{vogt11} Vogt, F., Martinache, F., Guyon, O., Yoshikawa, T., Yokochi, K., Garrel, V., \& Matsuo, T., 2011, \pasp, 123, 1434
\bibitem[Wang \it{et al.} (2014)]{wang14} Wang, J. J., \textit{et al.}, $2014$, Proc. of SPIE, 9147-195. 
\bibitem[Woillez \it{et al.} (2004)]{woillez04} Woillez, J. M., Perrin, G., Guerin, J., Lai, O., Reynaud, F., Wizinowich, P. L., Neyman, C. R., Le Mignant, D., Roth, K., \& White, J., 2004, Proc. of SPIE, 5491, 1425
\end{thebibliography}
\end{document}